\PassOptionsToPackage{dvipsnames}{xcolor}
\documentclass[twocolumn]{aastex63}

\def\cii      {[C\thinspace {\scriptsize II}]}

\usepackage{natbib}

\received{August 9, 2021}
\revised{December 10, 2021}
\accepted{December 28, 2021}

\submitjournal{ApJ}

\shorttitle{200\,pc imaging of a $z\,\sim\,7$ quasar host galaxy.}
\shortauthors{Walter et al.}

\graphicspath{{./}{figures/}}

\begin{document}

\title{ALMA 200\,pc imaging of a $z\,\sim\,7$ quasar reveals
a compact, disk--like host galaxy}

\correspondingauthor{Fabian Walter}
\email{walter@mpia.de}

\author[0000-0003-4793-7880]{Fabian Walter}
\affil{Max Planck Institute for Astronomy, K\"onigstuhl 17, 69117 Heidelberg, Germany}
\affil{National Radio Astronomy Observatory, Pete V. Domenici Array Science Center, P.O. Box O, Socorro, NM 87801, USA}

\author[0000-0002-9838-8191]{Marcel Neeleman}
\affil{Max Planck Institute for Astronomy, K\"onigstuhl 17, 69117 Heidelberg, Germany}

\author[0000-0002-2662-8803]{Roberto Decarli}
\affil{INAF—Osservatorio di Astrofisica e Scienza dello Spazio, via Gobetti 93/3, I-40129, Bologna, Italy}

\author[0000-0001-9024-8322]{Bram Venemans}
\affil{Leiden Observatory, Leiden University, PO Box 9513, 2300 RA Leiden, The Netherlands}
\affil{Max Planck Institute for Astronomy, K\"onigstuhl 17, 69117 Heidelberg, Germany}

\author[0000-0001-5492-4522]{Romain Meyer}
\affil{Max Planck Institute for Astronomy, K\"onigstuhl 17, 69117 Heidelberg, Germany}

\author[0000-0003-4678-3939]{Axel Weiss}
\affil{Max-Planck-Institut f\"ur Radioastronomie, Auf dem H\"ugel 69, 53121 Bonn, Germany}

\author[0000-0002-2931-7824]{Eduardo Ba\~nados}
\affil{Max Planck Institute for Astronomy, K\"onigstuhl 17, 69117 Heidelberg, Germany}

\author[0000-0001-8582-7012]{Sarah E.~I.\ Bosman}
\affil{Max Planck Institute for Astronomy, K\"onigstuhl 17, 69117 Heidelberg, Germany}

\author[0000-0001-6647-3861]{Chris Carilli}
\affil{National Radio Astronomy Observatory, Pete V. Domenici Array Science Center, P.O. Box O, Socorro, NM 87801, USA}

\author[0000-0003-3310-0131]{Xiaohui Fan}
\affil{Steward Observatory, University of Arizona, 933 North Cherry Avenue, Tucson, AZ 85721, USA}

\author[0000-0001-9585-1462]{Dominik Riechers}
\affil{Cornell University, 220 Space Sciences Building, Ithaca, NY 14853, USA}

\author[0000-0003-4996-9069]{Hans--Walter Rix}
\affil{Max Planck Institute for Astronomy, K\"onigstuhl 17, 69117 Heidelberg, Germany}

\author[0000-0003-2377-9574]{Todd A. Thompson}
\affil{Department of Astronomy and Center for Cosmology and Astro-Particle Physics, 140 W. 18th Ave, Columbus, Ohio 43210, USA}

\begin{abstract}

We present 0$\farcs$035 resolution ($\sim$200\,pc) imaging of the 158$\mu$m \cii{} line and the underlying dust continuum of the $z\,=\,6.9$ quasar J234833.34–305410.0. The 18\,h ALMA observations reveal extremely compact emission (diameter\,$\sim$\,1\,kpc) that is consistent with a simple, almost face--on, rotation--supported disk with a significant velocity dispersion of $\sim$\,160\,km\,s$^{-1}$. The gas mass in just the central 200\,pc is $\sim$\,4\,$\times$\,10$^{9}$\,M$_\odot$, about a factor two higher than that of the central supermassive black hole. Consequently we do not resolve the black hole's sphere of influence, and find no kinematic signature of the central supermassive black hole. Kinematic modeling of the \cii{} line shows that the dynamical mass at large radii is consistent with the gas mass, leaving little room for a significant mass contribution by stars and/or dark matter. The Toomre--Q parameter is less than unity throughout the disk, and thus is conducive to star formation, consistent with the high infrared luminosity of the system. The dust in the central region is optically thick, at a temperature $>$\,132\,K. Using standard scaling relations of dust heating by star formation, this implies an unprecedented high star formation rate density of $>$\,10$^4$\,M$_\odot$\,yr$^{-1}$\,kpc$^{-2}$. Such a high number can still be explained with the Eddington limit for star formation under certain assumptions, but could also imply that the central supermassive black hole contributes to the heating of the dust in the central 110\,pc.

\end{abstract}

\keywords{galaxies: high-redshift; galaxies: ISM; quasars: emission lines; quasars: general, quasar: individual: J234833.34--305410.0}

\section{Introduction} 
\label{sec:intro}

Since their discovery in the early 2000’s \citep[e.g.,][]{fan00}, the existence of luminous quasars at $z\,>\,6$, when the Universe was less than one billion years old, has been a puzzle. Optical/near-infrared (rest--frame UV) spectroscopy of these quasars revealed the typical signatures of broad emission line regions (BLRs, with linewidths of many 1000s of km\,s$^{-1}$) on top of continuum emission from the accretion disk. These broad emission lines are thought to emerge from a region that is very close to the central accreting supermassive black hole ($\ll\,1\,$pc), and they thus provide unique probes of the properties of the central source. If local scaling relations relating broad line features to black hole masses are employed, the BLR signatures point towards black hole masses exceeding a billion solar masses in many cases, putting strong constraints on early supermassive black hole growth \citep{wu15, mazzucchelli17, banados18, yang20, wang21}.

Over the last decade, (sub--)millimeter telescopes such as the Plateau de Bure Interferometer / Northern Extended Millimeter Array (NOEMA) and the Atacama Large Millimeter/sub-millimeter Array (ALMA) have provided the first constraints on the galaxies that host these central accreting black holes, in particular through spatially--resolved observations of the 158 $\mu$m \cii{} emission line and the underlying dust continuum. Overall, these studies show that the interstellar medium in the quasar host galaxies is rather compact with a typical extent that does not exceed a few kiloparsecs \citep[e.g.,][]{walter09,wang13,shao17,venemans20,novak20}, with a merger fraction of about $\sim$30\% \citep{neeleman19,neeleman21}.
Comparing the gas reservoirs with GAIA--corrected positions of the central accreting black hole indicates that the black holes are indeed located in the centers of the quasar host galaxies \citep{venemans20}. Kinematical analyses of the central regions of quasar host galaxies based on \cii{} observations point at dynamical masses that are $<10^{11}$\,M$_\odot$ \citep{walter09,shao17,pensabene20,izumi21a,neeleman21,yue21}, with gas masses contributing a significant fraction of the total dynamical masses. These results lead to the emerging picture that these early supermassive black holes reside in rapidly assembling galaxies that are quickly building up their stellar mass via both mergers and intense star formation.

The unprecedented angular resolution of ALMA now enables studies of the interstellar medium (ISM) in quasar host galaxies down to a few hundred parsec scales, rivaling the spatial resolution obtained in ISM surveys of nearby galaxies \citep{walter07,leroy09}. The quasar J234833.34--305410.0 at $z\,=\,6.9$ \citep[hereafter J2348--3054, ][]{venemans13} is a luminous broad absorption line (BAL) quasar powered by a black hole with a mass of $2.1 \times 10^9\,$M$_\odot$ \citep{derosa14, mazzucchelli17}, and is one of the dozen $z$\,$\sim$\,7 quasars currently known. Early spatially--unresolved ALMA observations (resolution: $0\farcs74\,\times\,0\farcs54$) targeting the \cii{} emission line revealed a highly significant \cii{} detection with a line width  FWHM\,=\,405$\pm$69\,km\,s$^{-1}$ and a luminosity of $L_{\rm [CII]}\,=(1.9 \pm 0.3) \times 10^9$\,L$_\odot$, as well as an underlying continuum flux density of $f_c\,=\,1.92 \pm 0.14$\,mJy \citep{venemans16}. These observations also revised J2348--3054's original redshift of $z\,=\,6.889\pm0.007$ from the \ion{Mg}{2} measurement \citep{derosa14} to a more accurate value of $z\,=\,6.9018\pm0.0007$ using the \cii{} line \citep{venemans16}. Follow--up observations at higher spatial resolution ($\sim\,0\farcs16$) yielded consistent line and continuum fluxes (\cii{} flux of 1.53$\pm$0.16\,Jy\,km\,s$^{-1}$, a \cii{} line width of 457$\pm$49\,km\,s$^{-1}$ and an underlying continuum of 2.28$\pm$0.07\,mJy), but still did not spatially resolve the emission significantly \citep{venemans20}. This situation, i.e.\ bright and centrally concentrated \cii\ emission, makes J2348--3054 a unique target to probe the kinematics in the vicinity of the central supermassive black hole using even higher resolution observations with ALMA. 

\begin{figure*}
\begin{center}
\includegraphics[width=0.75\textwidth]{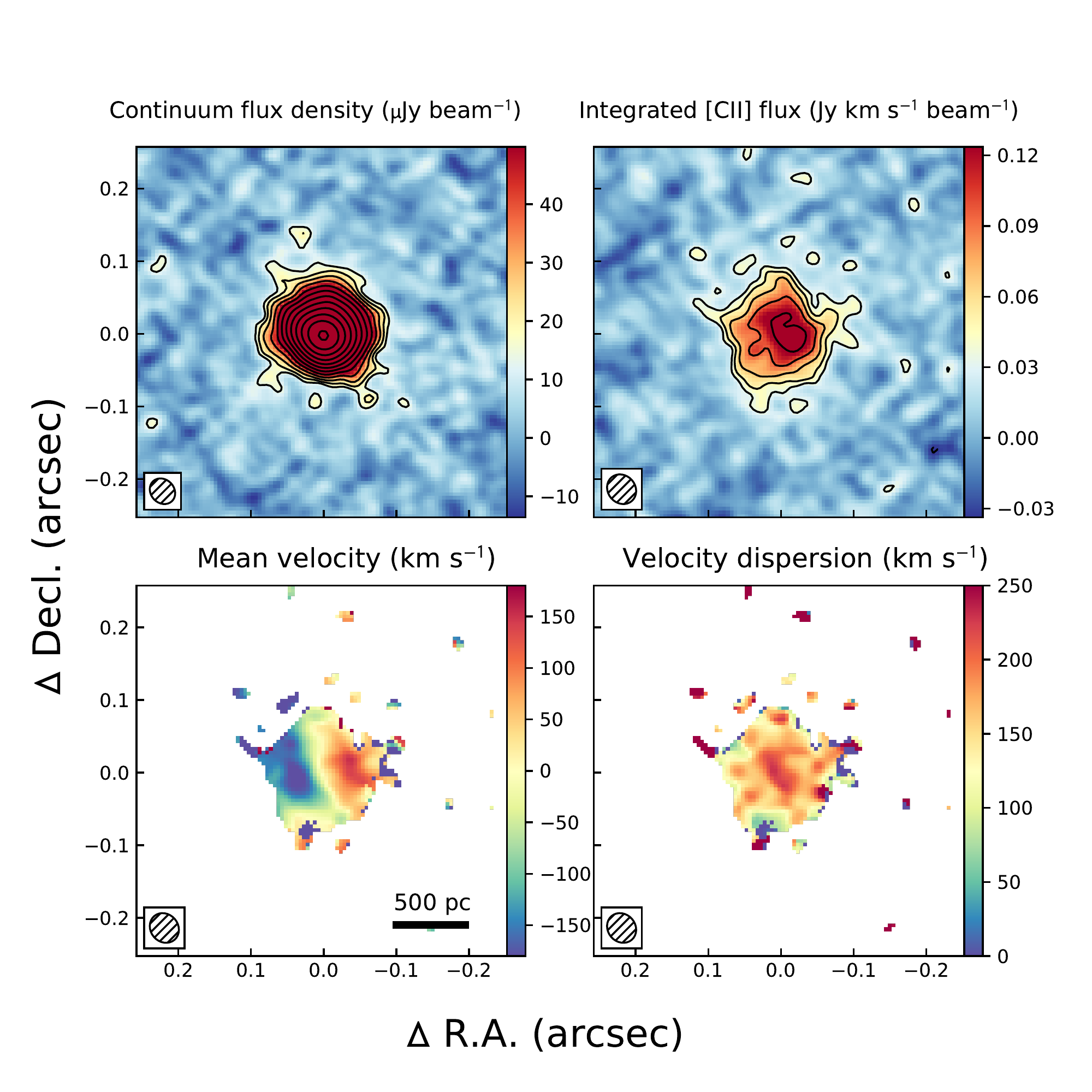}
\end{center}
\caption{\emph{Top left:} Restframe $\sim$\,1900\,GHz continuum map of J2348--3054. \emph{Top right:} Integrated continuum--subtracted \cii{} emission line. In the upper panels, contours are given at $\pm$ 3\,$\sigma$ and increase in powers of $\sqrt{2}$. \emph{Bottom left:} Mean velocity of the \cii\ emission line. \emph{Bottom right:} Velocity dispersion of the \cii\ emission. The bottom quantities are estimated from fitting Gaussian spectral profiles to each individual pixel. In the lower panels, colors indicate the velocities in units of km\,s$^{-1}$ (see color bars). The beam is shown as an inset in all panels.}
\label{fig:moments}
\end{figure*}

We here present $\sim$200\,pc resolution \cii\ and underlying dust continuum imaging of J2348--3054, pushing the capabilities of ALMA. In Sec.~\ref{sec:observations} we describe the ALMA observations. In Sec.~\ref{results} we summarize and analyze our observational results. This is followed by kinematic modeling of the \cii{} emission line in Sec.~\ref{sec:kinematics}. 
We present our conclusions in Sec.~\ref{summary}. Throughout this paper we use cosmological parameters H$_0$\,=\,70 km\,s$^{-1}$ Mpc$^{-1}$, $\Omega_{\rm M}$\,=\,0.3, and $\Omega_{\rm \Lambda}$\,=\,0.7, in agreement with \citet{planck16}, leading to a scale of 5.27\,kpc per arcsec at $z\,=\,6.9$.

\section{Observations and Methods}
\label{sec:observations}

Observations of J2348--3054 were obtained with ALMA in configuration C43--8 between 2019 July 9 and 19 for a total of 18.2\,h (9.0\,h on--source). These observations targeted the \cii{} line as well as the underlying dust continuum emission at $\simeq$\,240.5\,GHz in the lower sideband, and continuum--only emission in the upper sideband at $\sim$249.8\,GHz. The quasar J2258--2758 was used for flux and bandpass calibration and the quasar J2353--3037 was observed for phase calibration. These new high--resolution observations contained sufficient short spacings to recover the total flux (as detailed below, Sec.~\ref{sec:tot_flux}), and therefore they were not combined with the earlier, lower--resolution ALMA observations, which would have given too much weight to short baselines. 

The data presented in this paper were weighted using a robust weighting scheme resulting in a synthesized beam with major axis $a$\,=\,$0\farcs039$, minor axis $b$\,=\,$0\farcs032$, and a corresponding beam area of $\pi/(4\,\ln (2)) \times a \times b$ = 0.0014 arcsec$^2$. This corresponds to an effective radius of $r$=0.021$\arcsec$, or 110\,pc at the redshift of J2348--3054. The continuum emission was subtracted from the data cube using a first order polynomial fit in the UV plane by selecting the channels in the sideband covering the \cii{} emission that did not contain line emission. The noise in a 31.2\,MHz ($\sim$39\,km\,s$^{-1}$)--wide channel is 53\,$\mu$Jy\,beam$^{-1}$, and the data cube was cleaned down to a level of 2\,$\sigma$, using a central clean region, which was a circle of 0\farcs5 radius. A continuum map was created from the channels that did not contain line emission (using the three remaining spectral windows), resulting in a root mean square (rms) noise in the continuum map of 4.6\,$\mu$Jy\,beam$^{-1}$.

\begin{figure}
\begin{center}
\includegraphics[width=6cm,angle=270]{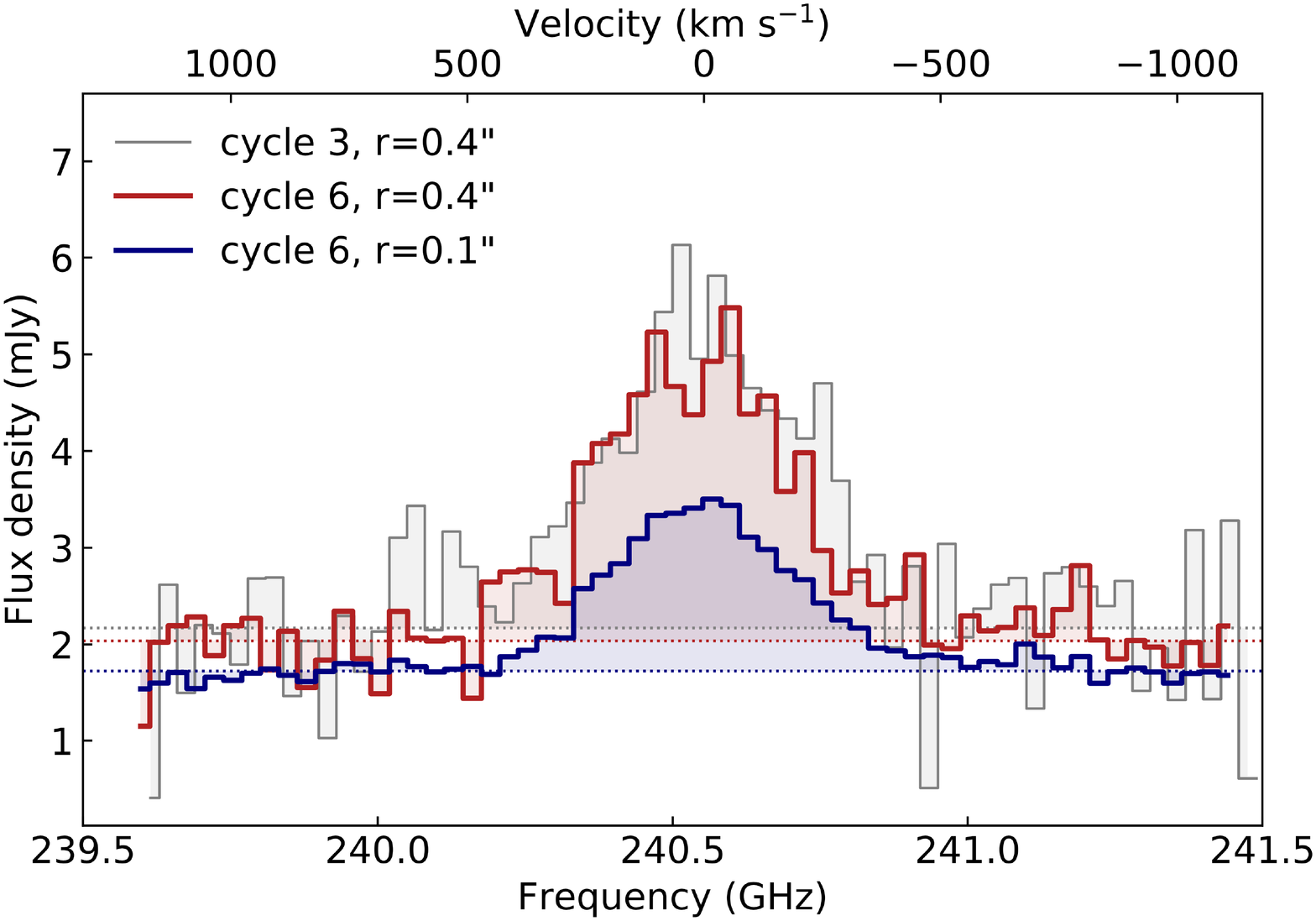}
\end{center}
\caption{J2348--3054 ALMA spectra of the high--resolution data discussed here (labeled `cycle~6') compared to the previous, unresolved measurements discussed in \citet{venemans16} (labeled `cycle~3'). For the new data two different apertures are shown (see discussion in Secs.~\ref{sec:tot_flux} and~\ref{sec:aperture}). We find that the new data fully recovers the flux seen in earlier low--resolution studies. The central 0$\farcs$1 region, corresponding to a radius of 0.53\,kpc, covers 89\% of the dust continuum emission and 50\% of the \cii{} line emission (see Tab.~\ref{tab:fluxes} for details).}
\label{fig:spectrum}
\end{figure}

\section{Resolved [CII] and dust emission}
\label{results}

\subsection{[CII] Moment and continuum maps}

In Fig.~\ref{fig:moments} we show the continuum map, the \cii{} intensity map, as well as the \cii{} velocity field and velocity dispersion maps. The latter maps were calculated by Gaussian fitting of the spectra at each position. We get very similar results for the velocity field and velocity dispersion maps if we calculate moment maps following the mathematical definitions, after clipping the emission at 2$\,\sigma$ in each channel \citep[see Appendix C in][]{neeleman21}. 
For reference, the \cii{} channel maps are presented in Appendix~\ref{sec:channels}. As we will see in Sec.~\ref{sec:kinematics}, the central peak in the velocity dispersion can be explained by beam smearing.

\subsection{Total [CII] Flux / Continuum Flux Density}
\label{sec:tot_flux}

In Fig.~\ref{fig:spectrum} we show the \cii{} spectrum in red, which includes the underlying continuum of J2348--3054, extracted over a circular aperture with a radius of 0.4$"$ (2.1\,kpc), encompassing the entire emission seen in the new observations. This spectrum was derived using the methodology outlined in \citet{jorsater95, walter99, walter08, novak19, novak20}, to account for the fact that the synthesized and clean beam areas in interferometric imaging have different integrals. In this $r$=0.4$"$ (2.1\,kpc) aperture, we derive a continuum flux of 2.00\,$\pm$\,0.07\,mJy, and a flux for the \cii{} line of 1.62\,$\pm$\,0.18\,Jy\,km\,s$^{-1}$ (linewidth: 481$\pm$71 km\,s$^{-1}$, $L_{\rm [CII]}$=1.8$\times10^9$ L$_\odot$), both in good agreement with the values reported in the earlier low--resolution observations \citep{venemans16}, implying that there is no significant emission outside the 2.1\,kpc aperture. We adopt these measurements as the total line and continuum fluxes of J2348--3054 and report them in Tab.~\ref{tab:fluxes}.

\subsection{Dust Temperature and Optical Thickness}
\label{sec:tau}

Typically, one proceeds calculating dust gas masses and star formation rates from the dust continuum measurements by assuming a temperature $T_{\rm d}$ and emissivity index $\beta$,  and optically thin emission \citep{dunne2000,dunne2001,beelen06}.

As we will see below, the flux densities per surface area in J2348--3054 are so extreme, that we approach optically thick emission at our resolution. We here thus proceed using the full radiative transfer equation to relate the observed flux densities to the intrinsic properties, following:
\begin{equation}
S_\nu\,=\,\Omega_{\rm a}\,[B_\nu(T_d)-B_\nu(T_{\rm CMB})]\,[1-\exp(-\tau_\nu)]\,(1+z)^{-3},
\label{eq:snu}
\end{equation}
where $S_\nu$ is the dust continuum flux density measured at $\nu$=1900.54\,GHz (the rest frequency of the \cii{} emission), $\Omega_{\rm a}$ is the solid angle corresponding to our aperture in steradians, $B_\nu(T_d)$ and $B_\nu$($T_{\rm CMB}$) are the black body emission ($B_\nu(T)=2\,h\,\nu^3\,c^{-2} [\exp(h\,\nu/(k_{\rm b} T)) -1]^{-1}$) from the dust and cosmic microwave background (CMB), respectively, and $\tau_\nu$ is the frequency--dependent optical depth of the dust \citep[see, e.g.,][]{draine03,weiss07}. From this equation we see that for a given redshift $z$, flux density $S_\nu$, solid angle $\Omega_{\rm a}$, and a given optical depth $\tau_\nu$, the dust temperature is uniquely determined. This temperature is beam--averaged, and will be a lower limit for filling factors $\eta\,<\,1$ within the aperture $\Omega_{\rm a}$.

We note that $\tau_\nu$ is related to the total dust mass, $M_{\rm dust}$, via, $\tau_\nu=\kappa_0 (\nu / \nu_{\rm ref})^\beta\,M_{\rm dust}\,A^{-1}$, where $\kappa_0=13.9$\,cm$^2$\,g$^{-1}$ is the absorption coefficient per unit dust mass, $\beta$ is the emissivity index, $\nu_{\rm ref}=2141$\,GHz is the reference frequency for the dust emissivity \citep{draine03}, $M_{\rm dust}$ is the dust mass and $A=\pi r^2$ is the area of the emitting region.

\subsection{Total emission (r = 2.1 kpc)}
\label{sec:total}

In Appendix~\ref{app:band_8} we present ALMA Compact Array (ACA) band 8 continuum data, as well as archival WISE and Herschel data, which put constraints on the dust spectral energy distribution (SED) of J2348--3054. From this, we derive a dust temperature of $T_{\rm d}\,=\,84.7^{+8.9}_{-10.5}$\,K,  an emissivity index of $\beta\,=\,1.21^{+0.20}_{-0.15}$
and an integrated total infrared luminosity of $L_{\rm TIR}$=3.2\,$\times$\,10$^{13}$\,L$_\odot$. This gives a total dust mass of $M_{\rm dust}\,\simeq\,1.1^{+0.41}_{-0.25}\times10^{8}$\,M$_\odot$. Assuming a gas--to--dust ratio of 100 \citep[e.g.,][]{berta16}, this implies a total molecular gas mass of $M_{\rm H2}\,\simeq\,1.1\times10^{10}$\,M$_\odot$. If we instead use the \cii{} emission as a tracer for the molecular gas, following, e.g., \citet{zanella18}, we derive $M_{\rm H2, [CII]}$\,=\,$5.4\times10^{10}$\,M$_\odot$. However, this conversion likely over--predicts the molecular mass estimates in quasar host galaxies \citep{neeleman21}. We note that \citet{venemans17} derived a molecular gas mass of $M_{\rm H2}$\,=\,$1.2\times10^{10}$\,M$_\odot$ based on CO(6--5) and CO(7--6) observations. We conclude that our dust--based H$_2$ mass measurement is in broad agreement with those numbers. 

From the total infrared luminosity we derive SFR$_{\rm TIR}$ of 4700\,M$_\odot$\,yr$^{-1}$ using the relation in \citet{kennicutt12}. We can also estimate the SFR based on the \cii{} line, following, e.g., \citet[][their `high $\Sigma_{\rm TIR}$' relation]{herreracamus18}, and derive a \cii{}--based SFR$_{\rm [CII]}$ of 530\,M$_\odot$\,yr$^{-1}$. Some of the difference can be attributed to the  well--known \cii{} `deficit'  (Sec.~\ref{sec:deficit}). Since we do not have spatially resolved information on the dust SED available for the other wavelengths, we continue the discussion and analysis that follows based on the high--resolution band~6 ALMA data only.

\begin{figure*}
\begin{center}
\includegraphics[width=7cm]{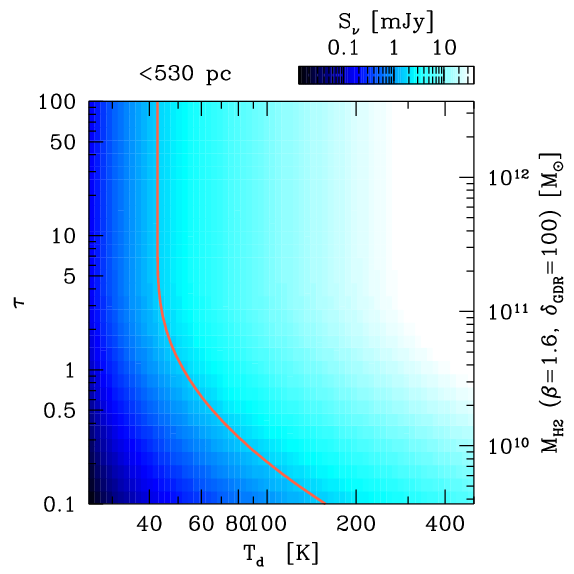}
\includegraphics[width=7cm]{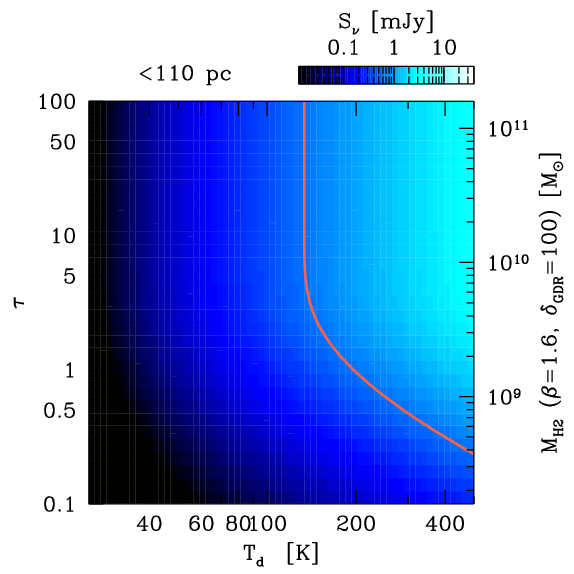}
\end{center}
\caption{Predicted dust--continuum flux density ($S_\nu$) at the \cii{} frequency as a function of the dust temperature $T_{\rm d}$ and optical depth $\tau$, following Eq.~\ref{eq:snu}, computed for two different apertures ({\em left:} $r\,=\,530$\,pc, {\em right:} $r\,=\,110$\,pc). Optical depths are converted into molecular gas mass within the aperture as described in Sec.~\ref{sec:tau}. The observed flux densities are marked with red lines. For a given flux density, $T_{\rm d}$ and $\tau$ (and $M_{\rm H2}$) are anti-correlated in the optically thin regime ($\tau\ll 1$). On the other hand, in the optically-thick regime $T_{\rm d}$ asymptotically approaches a minimum value as $\tau$ increases. The minimum temperature is $42.4$\,K for the 530\,pc aperture, and it increases to 132\,K for the central 110\,pc. Optical depth of unity is reached with a gas mass of $M_{\rm H2}=3.7\times10^{10}$\,M$_\odot$ ($1.6\times10^9$\,M$_\odot$) for the 530\,pc (110\,pc) aperture.
}
\label{fig:tau_530}
\end{figure*}

\subsection{0.1" (r = 530 pc) aperture}
\label{sec:aperture}
In Fig.~\ref{fig:spectrum} we also show the spectrum extracted over a central aperture with a radius of 0\farcs1 (530\,pc) as a blue line. This aperture size was chosen to encompass all  emission that is visible in the integrated \cii{} map (Fig.~\ref{fig:moments}) above $2\sigma$. We measure a continuum flux density $\sim$10\% lower than the total, whereas the \cii{} emission is decreased by $\sim$50\% (fluxes reported in Tab.~\ref{tab:fluxes}, column 3). We note that the \cii{} line widths are the same within uncertainties --- we are not picking up higher velocity gas when changing the aperture. From this simple comparison, we can already conclude that the \cii{} is not as centrally concentrated as the dust continuum emission. In Fig.~\ref{fig:tau_530} (left) we plot, for a given value of $\tau$ and T$_{\rm dust}$ the resulting flux density S$_{\nu}$ of the 530\,pc aperture ($\Omega_{\rm a}=\pi\times$0\farcs1$\times$0\farcs1=$7.4\times10^{-13}$\,sr). We also plot, as a red line, the measured flux density of this aperture (S$_{\nu}$=1.77\,$\pm$\,0.01\,mJy; see Tab.~\ref{tab:fluxes}). From this plot we deduce that the dust temperature must be at least 42.4\,K assuming optically thick emission (consistent with the temperature derived in Sec.~\ref{sec:total}). For a temperature of 84.7\,K (Sec.~\ref{sec:total}) we derive an optical depth of $\tau$\,=\,0.262.

\subsection{The central resolution element (r = 110 pc)}
\label{sec:beam}
We now concentrate on the central resolution element (central beam) of the observations, which, for an effective radius of 0.021$"$ or 110\,pc (Sec.~\ref{sec:observations}) corresponds to an area of 0.039\,kpc$^2$ (solid angle $\Omega_{\rm a}=\pi\times $0\farcs021$\times$0$\farcs$021=$3.3\times10^{-14}$\,sr). In this central beam, we derive a flux density of 0.64\,$\pm$\,0.01\,mJy in the continuum and a \cii\ line flux of 0.11\,$\pm$\,0.01\,Jy\,km\,s$^{-1}$. From Fig.~\ref{fig:tau_530} (right, same as left hand plot but for the central resolution element) we find that temperatures $T_{\rm d}<$132\,K (optically thick case) are ruled out by our measurement. If we assume an optical thickness as high as $\tau$\,$\sim$\,4, we derive a total molecular gas mass of 6\,$\times$\,10$^{9}$\,M$_\odot$ (with a corresponding temperature of 132\,K). For $\tau$\,$=$\,1, we derive a total molecular gas mass of 1.6\,$\times$\,10$^{9}$\,M$_\odot$ (with a corresponding temperature of 183\,K).
We thus adopt an H$_2$ mass of  $M_{\rm H2}$=$(4\,\pm\,2)\times 10^{9}$\,M$_\odot$ for the central resolution element and note that it exceeds the mass of the central supermassive black hole. The resulting average H$_2$ mass surface density is $\Sigma_{\rm H2}$\,=\,$(10\,\pm\,5)\times 10^{4}$\,M$_\odot$\,pc$^{-2}$. 

For a temperature of 132\,K (183\,K) we derive a total infrared luminosity $L_{\rm TIR}=6.5\times10^{12}$\,L$_\odot$ ($2.3\times10^{13}$\,L$_\odot$) and, assuming that the dust is heated by star formation \citep{kennicutt12}, a SFR of 970\,M$_\odot$\,yr$^{-1}$ (3,600\,M$_\odot$\,yr$^{-1}$) for the central $r$\,=\,110\,pc beam. Proceeding with the lower temperature/$L_{\rm TIR}$ this corresponds to a SFR surface density of $\Sigma_{\rm SFR}\,\sim\,25.500$\,M$_\odot$\,yr$^{-1}$\,kpc$^{-2}$ (averaged over the central beam). This very high star formation rate surface density is due to the very high dust temperature implied by our measurement. However we also note that we cannot rule out some contribution to the dust heating by the central accreting supermassive black hole.

Taking this exceptionally high star formation rate surface density at face value, it is interesting to compare the observed luminosity for J2348--3054 with the Eddington limit for dusty gas \citep{thompson05}. For a geometrically thin disk, the Eddington flux is
\begin{eqnarray}
    F_{\rm Edd} & = & \frac{2\pi G \Sigma_{\rm tot} c}{\kappa_R} \nonumber \\ 
    & \simeq & 1.3\times10^{14}\,{L_\odot\,\,\rm kpc^{-2}}\left(\frac{\Sigma_{\rm tot}}{10^5\,M_\odot\,\,\rm pc^{-2}}\right)\left(\frac{5\,\rm cm^2\,\,g^{-1}}{\kappa_R}\right), \nonumber \\ 
    \label{eddington_flux}
\end{eqnarray}
where $\Sigma_{\rm tot}$ is the total mass surface density and $\kappa_R$ is the temperature-dependent Rosseland-mean opacity, which for midplane temperatures above $\gtrsim200$\,K is approximately constant at $\kappa_R\sim5-10$\,cm$^2$ per gram of gas, assuming a Milky Way-like dust-to-gas ratio \citep[e.g.,][]{semenov03}. In the latter equality, we have scaled $\Sigma_{\rm tot}$ and $\kappa_R$ to values appropriate for the inner 110\,pc of J2348--3054. The total Eddington luminosity from both sides of the disk is $L_{\rm Edd}\simeq2\pi r^2F_{\rm Edd}\simeq1.0\times10^{13}$\,L$_\odot$. This value is in good agreement with the observed luminosity for the central region of J2348--3054, and suggest that dust may play a critical role for the dynamics in this central region \citep{thompson05,Krumholz2012,
Krumholz2013,Davis2014,Zhang2017}.

We note that the above Eddington flux is significantly larger than the  ``characteristic" values from \citet{thompson05} for radiation pressure supported starbursts ($F_{\rm Edd}\simeq10^{13}$\,L$_\odot$ kpc$^{-2}$). This is because those characteristic values were derived under the assumption that $\kappa_R\propto T^2$, which is valid for lower midplane temperatures  of $T_{\rm mid}\lesssim200$\,K \citep{semenov03}. To estimate the midplane temperature for J2348--3054, we use the Eddington flux in equation (\ref{eddington_flux}) to calculate an effective temperature of $T_{\rm eff}=(F_{\rm Edd}/\sigma_{\rm SB})^{1/4}\simeq170$\,K. For a simple plane-parallel atmosphere, the effective temperature and midplane temperature are related via $T_{\rm mid}^4\simeq(3/4)\tau_R T_{\rm eff}^4$, where $\tau_R=\kappa_R\Sigma_g/2$, and $\Sigma_g\simeq10^5$\,M$_\odot$ pc$^{-2}$ is the gas surface density. Using the observed values, this implies a mid--plane temperature of $T_{\rm mid}\simeq430$\,K for J2348--3054. This both justifies our assumption of a constant $\kappa_R$, and provides an explanation for the order of magnitude larger Eddington flux seen in J2348--3054 compared to the characteristic values in \citet{thompson05}.

\begin{deluxetable*}{lccc}
\label{tab:fluxes}
\tablecolumns{4}
\tablehead{
\colhead{} & \colhead{total} & \colhead{$r$\,=\,530\,pc aperture} & \colhead{central pixel (r\,$<$\,110\,pc) $^{a}$}   }
\startdata
$f_c$ (mJy)                    &     2.00\,$\pm$\,0.07   &     1.77\,$\pm$\,0.01         &         0.64\,$\pm$\,0.01  \\
\cii\ (Jy\,km\,s$^{-1}$)      &     1.62\,$\pm$\,0.18     &     0.86\,$\pm$\,0.03         &         0.11\,$\pm$\,0.01   
\enddata
\caption{Continuum flux density and \cii{} line flux for the entire host galaxy (second column), and a $r$\,=\,530\,pc and the central pixel (third and forth column). \\$^a$ The central pixel is defined as the brightest continuum pixel, which is slightly offset from the peak of the [CII] emission map (Fig.~\ref{fig:moments}).
}
\end{deluxetable*}

\subsection{Radial dependence of the infrared luminosity}
\label{sec:radial}

\begin{figure}
\begin{center}
\includegraphics[width=0.5\textwidth]{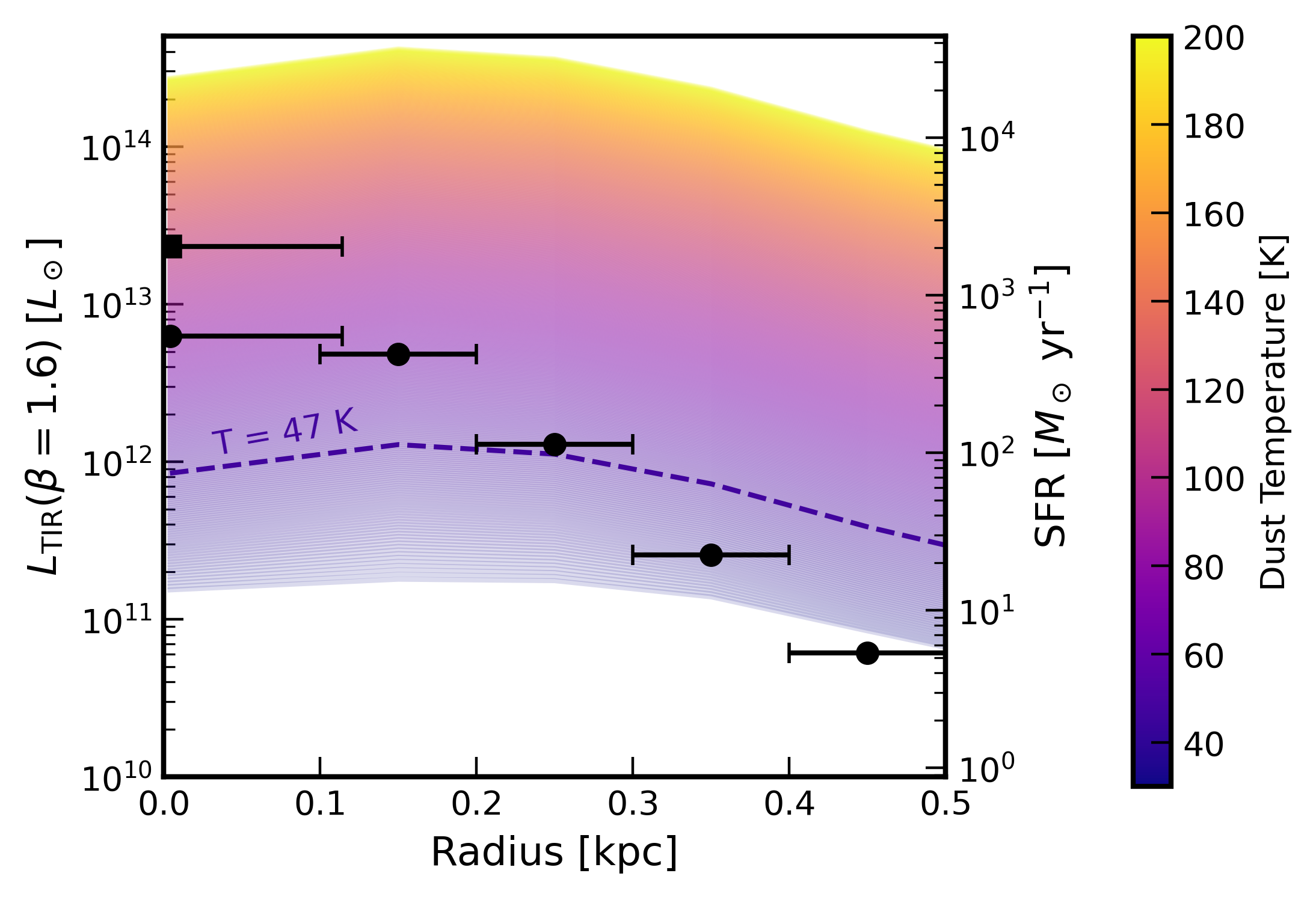}
\end{center}
\caption{Azimuthally averaged infrared luminosity $L_{\rm{TIR}}(8-1000 \mu m)$ for J2348--3054 in 0.1\,kpc--wide annuli from 0.1 to 0.5 kpc. The derived luminosities take into account the heating by, and contrast to, the CMB. The colored lines show the luminosities that would be derived using a constant dust temperature (tempertures given in color wedge to the right). The canonical value of T$_{\rm d}$\,=\,47\,K is shown as a dashed line. The measurements shown in black are calculated using the miminum (optically thick) temperatures derived using Eq.~\ref{eq:snu} (see text for details). For the central beam, we also show the resulting L$_{\rm TIR}$ for a temperature of 132 K (183 K) with a circle (square).
\label{fig:radial}}
\end{figure}

We now examine the inferred infrared luminosity profile and its dependence on the choice of the assumed dust temperature. This is shown in Fig.~\ref{fig:radial}, where we plot the azimuthally averaged infrared luminosity in bins of 0.1\,kpc. As the resulting luminosity is a function of temperature, we show a range luminosities based on (constant) temperatures as colors. We overplot the luminosities in each radial bins based on the minimum (optically thick) temperatures derived using Eq.~\ref{eq:snu}. For the radial bins [0--0.1, 0.1--0.2, 0.2--0.3, 0.3--0.4, 0.4--0.5]\,kpc the corresponding dust temperatures are [135, 99, 65, 43, 31]\,K. We note that towards large radii, the temperatures derived in the optically thick case are lower than the canonical T\,=\,47\,K value (see also Secs.~\ref{sec:total} and~\ref{sec:aperture}), whereas they exceed that value for the two central bins (Sec.~\ref{sec:beam}). Consequently, this leads to significantly increased infrared luminosities in the central region. 

We note however that even in the center the minimum temperature is significantly below the temperatures expected for central heating of the quasar \citep[where temperatures exceeding 1.000\,K are observed, e.g.][]{jiang06,jiang10}. 
Matched--resolution, multi--band ALMA continuum observations would be needed to constrain the dust temperature in J2348--3054 further and to disentangle the impact of the black hole / AGN torus heating from heating by stars.

\subsection{Spatial variations of the [CII] `deficit'}
\label{sec:deficit}

In local spiral galaxies, the luminosities of \cii{} and TIR scale roughly linearly, with a typical \cii{}/TIR luminosity ratio of $\sim$5$\times$10$^{-3}$. In regimes of very high far--infrared luminosity densities, however, the ratio substantially drops. This is typically referred to as the `\cii{} deficit' \citep[e.g.,][]{diazsantos17,smith17}. For the entire J2348--3054 system we derive a ratio of $L_{\rm [CII]}/L_{\rm TIR}=6.5 \times 10^{-5}$. This value is lower than typically found in other high--z quasar environments studied so far \citep[see, e.g.,][]{decarli18}. For the central 110\,pc region we derive an even lower number, $L_{\rm [CII]}$/$L_{\rm TIR}$=$2.6\times10^{-5}$.

This finding can be explained as follows: for a modified black body spectrum
the total infrared emission is related to the dust temperature with a power--law index that exceeds the Stefan--Boltzmann law\footnote{The reason for a power--law index $>4$ stems from Wien's displacement law: a low temperature spectrum is shifted to a lower opacity regime compared to warmer gas. This reduces the intensity of the low compared to the high temperature case, in addition to Stefan--Boltzmann's law.} (roughly L$_{\rm TIR}$\,$\propto$\,T$^{4.6}$ for the dust parameters in the central beam), while monochromatic line emission typically scales only linearly with increasing temperature. Thus, a strongly decreasing $L_{\rm [CII]}$/$L_{\rm TIR}$ ratio for warmer sources is expected.  

Given the high dust opacities in J2348--3054 in Sec.~\ref{sec:tau} there is, however, a secondary effect that reduces the expected \cii{} line emission: Ignoring the line opacity for now, the line intensity is set by the difference between the intensity for a given line excitation temperature and the background radiation field --- the latter increases for increasing dust opacities. To quantify this effect for J2348--3054, we built a simplistic model for the \cii{} emission using the observed \cii{} and dust continuum intensity profiles and employing RADEX.  For the central beam we assume a dust temperature of 190\,K and radially decreasing dust temperatures, such that the resulting dust column densities follow an exponential disk profile with a scale length of 200\,pc. We convert these dust column densities to \cii{} column densities using a fixed gas--to--dust mass ratio and a fixed \cii{} abundance relative to hydrogen. We further assume for simplicity collisional excitation by H$_2$, a constant density, and T$_{\rm kin}$=T$_{\rm dust}$ \footnote{The main purpose of the model is to investigate the effect of the background radiation field and not to describe the \cii{} excitation in a realistic manner. For simplicity we therefore here do not consider collisions with hydrogen and electrons.}. 

\begin{figure}
\begin{center}
\includegraphics[width=0.47\textwidth]{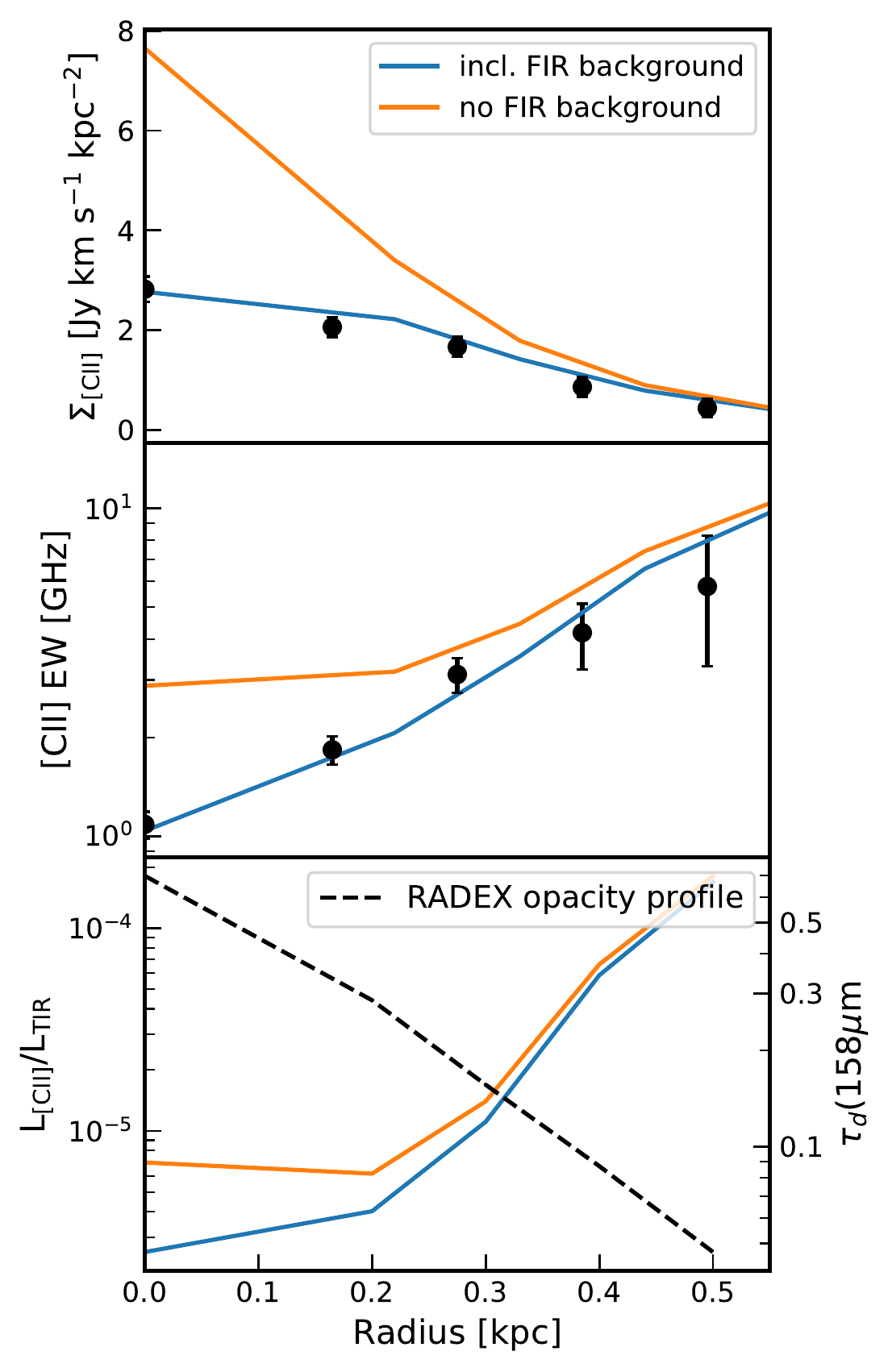}
\end{center}
\caption{RADEX--predicted radial \cii{} intensity profiles ({\em top}) with (blue) and without (orange) the background IR radiation field based on the model described in Sec.~\ref{sec:deficit}. The {\em middle} panel shows the resulting line over continuum ratios expressed as the line equivalent width for both cases. In the {\em bottom} panel we show the resulting $L_{\rm [CII]}$/$L_{\rm TIR}$ ratios, and the opacity profile in the RADEX model (dashed line, see text for details). The data points in the top and middle panel show the observed \cii{} surface brightness and \cii{} equivalent width, respectively. \label{fig:CIImodel}}
\end{figure}

We can thus determine the resulting \cii{} line intensities as a function of radius with and without considering an IR background field via RADEX (the CMB temperature is included as a background field in both cases).  The resulting radial \cii{} profiles are shown in Fig.~\ref{fig:CIImodel} (top) where we find a good match to the observed radial profile with an \cii{} abundance of 1\,$\times$\,10$^{-4}$ and a density of 1\,$\times$\,10$^{5}$\,cm$^{-3}$. The figure shows the strong impact of dust radiation field on the resulting \cii{} intensities. While the impact of the
radiation field is negligible in the outer parts of the disk (with low dust temperature and dust column density/opacity), the \cii{} intensity is reduced by a factor of $\sim$2.5 in the center of the disk. Integrating over the entire disk we find that the \cii{} intensity is reduced by a factor of 1.5 compared to the model without the IR background field. Fig.\,\ref{fig:CIImodel} middle shows the resulting \cii{} line--to--continuum ratios expressed as a function of \cii{} equivalent width. The bottom panel shows the resulting $L_{\rm [CII]}$/$L_{\rm TIR}$ ratios as a function of radius. The latter shows a strongly decreasing $L_{\rm [CII]}$/$L_{\rm TIR}$ ratio due to the increasing temperatures towards the center as the IR background field increases.

\section{Kinematics of the interstellar medium}
\label{sec:kinematics}

\subsection{Modeling the kinematics}
\label{sec:modeling}

\begin{figure}
\begin{center}
\includegraphics[width=0.5\textwidth]{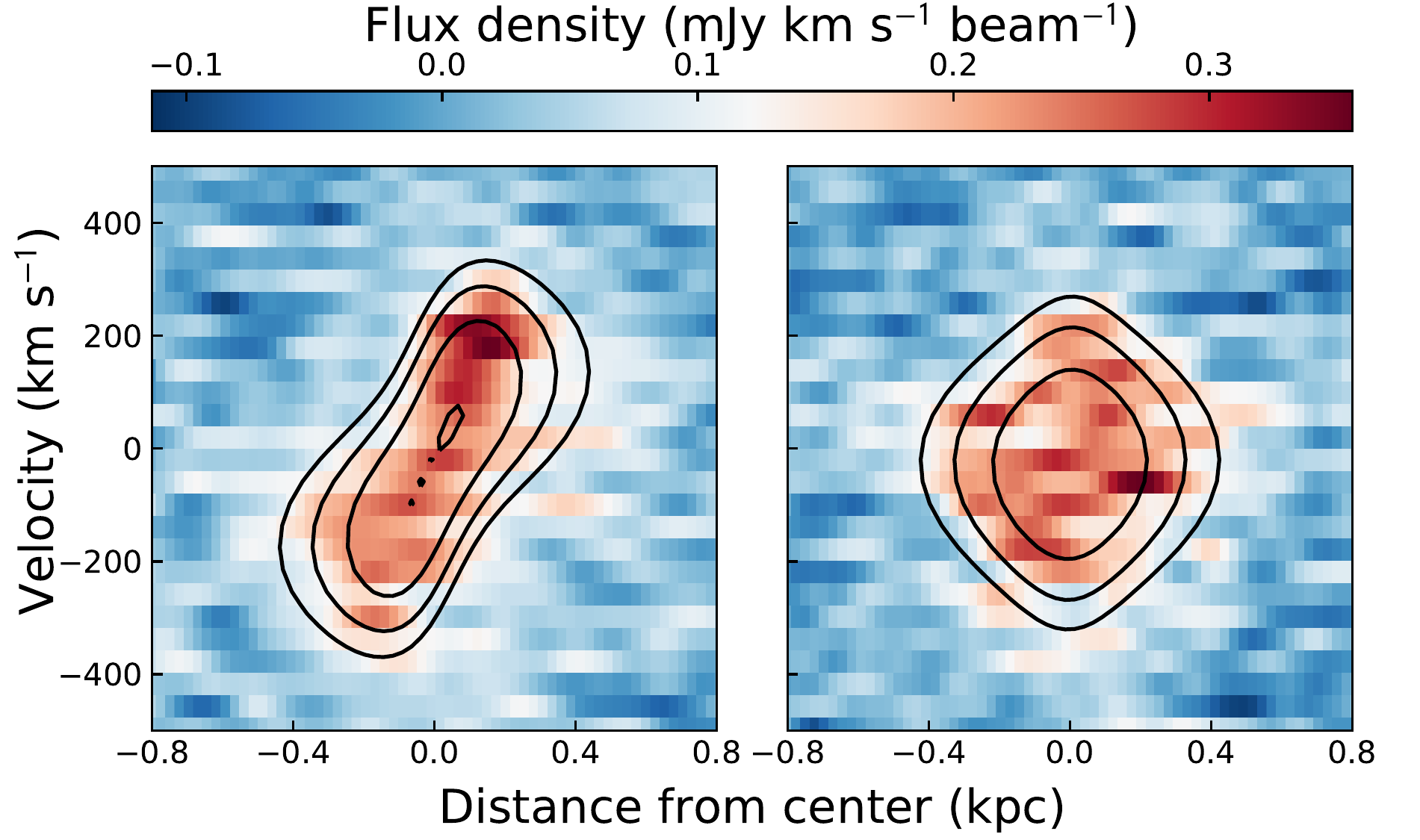}
\end{center}
\caption{
Position velocity diagrams ($p$--$v$ diagrams) along the major axis (\emph{left panel}) and minor axis (\emph{right panel}) of J2348--3054. The $p$--$v$ diagram for the infinitely thin disk model with constant rotational velocity is shown by the contours, which start at 2$\sigma$ and increase by powers of $\sqrt{2}$. We find that this simple model accurately describes the dynamics of the \cii\ emission line (see Appendix \ref{sec:model_pars} for additional models).}
\label{fig:pV}
\end{figure}

The \cii\ velocity field shown in Fig.~\ref{fig:moments} shows a clear gradient with a position angle of 275$\degr$. Such a velocity gradient is consistent with the emission arising from gas that is rotating. To model this gas, we assume that the \cii--emitting gas is constrained to a disk\footnote{ In Sec.~\ref{sec:scale_height} we will see that assuming a disk--like geometry is warranted.}. To estimate the kinematic parameters of the gas, i.e., rotational velocity and velocity dispersion, we have fitted the \cii\ emission line using the kinematic fitting code, \emph{Qubefit} \citep{neeleman21}. In short, \emph{Qubefit} uses a fully Bayesian approach to find the best--fit parameters to a user--defined model for the emission. In our case, we model the \cii\ emission using an infinitely thin disk. For the assumed rotation curve of the disk, we assume a constant velocity (i.e. a flat rotation curve) throughout the disk. We tested this assumption with both a linearly increasing velocity curve (solid body rotation) as well as an exponentially decreasing velocity curve, but found that neither curve provided an improved fit to the constant velocity case (see also Sec. \ref{sec:decomposition}, and Appendix~\ref{sec:model_pars}). The model fit along both the major and minor axis are given in Fig.~\ref{fig:pV}, and the results for the fitting are given in Tab.~\ref{tab:modelpar}. We find that this simple thin disk model can reproduce the observed \cii\ emission line remarkably well. In the residual channel maps (Fig.~\ref{fig:channels}), we see very little residual structure at $>$3$\sigma$; indicating that at this sensitivity the data can be accurately modeled with this disk model. This model also recovers the increased velocity dispersion seen in the central region, and since the velocity dispersion is assumed to be constant in the model, this increase is solely due to beam smearing effects.

With our fiducial constant velocity model, we find that the galaxy is nearly face-on with an inclination less than 25.7$\degr$ at the 98\,\% confidence level. This can also be seen in the integrated \cii\ emission from this galaxy in Fig.~\ref{fig:moments}, where the emission appears nearly circular.
The galaxy's velocity dispersion is $161\,\pm\,4$\,km\,s$^{-1}$, which is high, but consistent with the sample of $z\,>\,6$ quasars discussed in \citet{neeleman21}. The ratio of rotational velocity to velocity dispersion is a standard measure of the rotational support of a system \citep[e.g.,][]{epinat09, burkert10} with higher ratios indicating a greater level of rotational support. For J2348--3054 the ratio is $>$\,1.7, and the system is therefore likely rotationally supported, although highly turbulent.

\subsection{Rotation curve}
\label{sec:decomposition}

\begin{figure}
\begin{center}
\includegraphics[width=8.5cm]{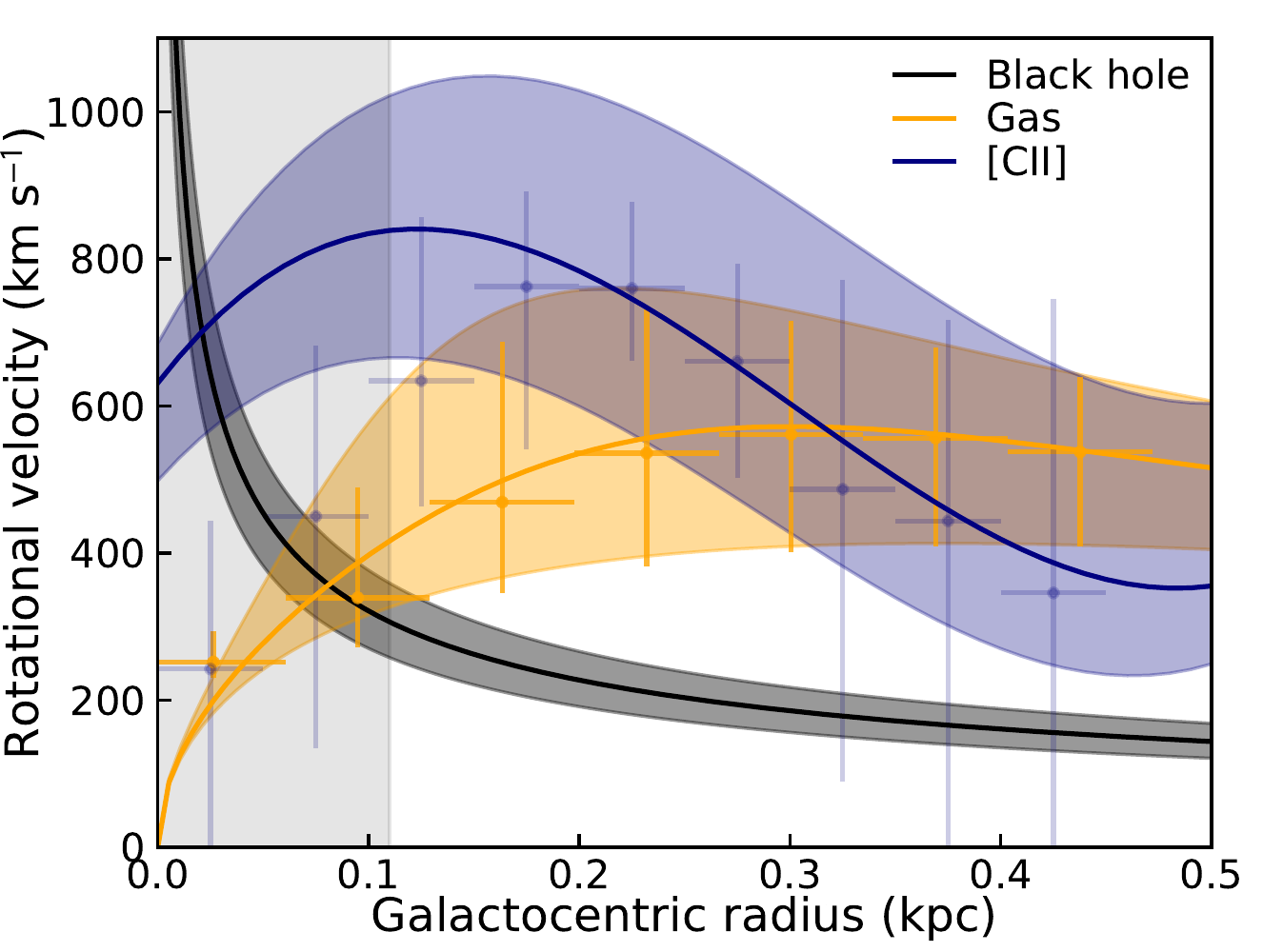}
\end{center}
\caption{Rotation curve of J2348--3054. The measurements in light violet show the rotation curve as derived from the velocity field in Fig.~\ref{fig:moments}; these points have not been corrected for the effect of the beam. To correct the rotation curve for this effect, we have used \emph{$^{3D}$Barolo} \citep{diteodoro2015}, assuming an inclination of 15$\degr$ (blue solid line). The violet--shaded region marks the 5$\degr$ uncertainty on the inclination parameter. The black solid curve is the Keplerian curve for a black hole of mass 2.1 $\times$ 10$^9$ M$_\odot$ where the dark gray--shaded region marks the uncertainty on this mass measurement. Finally, the orange measurement is the velocity contribution from the molecular gas (see text), where the line has been corrected for the effect of the beam, and the shaded region accounts for the large spread in temperatures at each radius (Sec.~\ref{sec:radial}). The effective size of the beam is marked by the gray--shaded region.
\label{fig:rotcurve}}
\end{figure}

To determine the dynamical importance of the individual mass constituents of the galaxy, we can determine their relative contribution to the rotation curve. This is shown in Fig.~\ref{fig:rotcurve}. To measure the total rotation curve, we use the kinematic information of the \cii\ line. Assuming that the velocity field of the \cii\ emission line is solely due to circular motion within the plane, we can convert the line-of-sight velocity measurements of Fig.~\ref{fig:moments} into an inclination--corrected rotation velocity \citep[see e.g.,][]{neeleman20}. The radially averaged measurements are shown by the blue points in Fig.~\ref{fig:rotcurve}, where the horizontal uncertainties denote the size of the radial bins, and the vertical uncertainties the 1$\sigma$ spread in the data in that bin. These values have not been corrected for the effect of the beam. To correct for this effect, we model the data using the code \emph{3DBarolo} \citep{diteodoro2015}. During the modeling, we fix the inclination at 15$\degr$ (solid blue line); the blue--shaded region marks the $\pm$5$\degr$ uncertainty on this inclination measurement. We see that at small radii the effect of the beam causes the data to underestimate the true rotation curve.

To get the velocity contribution of the black hole, we take the mass of the black hole and assume a simple Keplerian rotation curve. This is shown by the solid black line for a black hole with a mass of $2.1 \times 10^9\,$M$_\odot$ (Sec.~\ref{sec:intro}). Comparing this black curve with the rotation curve determined from the \cii\ line, we see that despite this large central black hole mass, the contribution of the black hole to the rotation curve is negligible for all radii. Only well within the current beam (gray--shaded region) does the rotation curve of the black hole become comparable to the observed rotation curve.

To measure the molecular gas contribution to the rotation curve, we take the dust continuum observations, and assume that the molecular mass is traced by this emission. To estimate the mass profile, we convert the infrared luminosity in each radial bin as defined in Sec.~\ref{sec:radial} to a dust mass estimate assuming the minimum, optically thick temperatures as derived in this section. We further assume a constant dust--to--gas ratio of 100 to convert the dust mass into a gas mass. We then assume the gas is distributed spherically (i.e. M$_{\rm dyn}$\,=\,r\,v$_{\rm rot}^2$/G, where G is the gravitational constant) to estimate the rotation curve from the molecular gas \citep[this underestimates the true rotation curve by at most 30\,\%, see e.g.,][]{walter97}. These measurements are shown by the orange data points where the vertical uncertainties account for different possible temperature gradients and dust--to--gas ratios. To correct these measurements for the effect of the beam, we fit an exponential model convolved with the beam to the continuum data to get an estimate of the true continuum flux distribution. Converting this flux measurement using the above mentioned approach yields the line shown in orange in Fig.~\ref{fig:rotcurve}. We can see that the contribution of the molecular gas is significantly larger than that of the black hole at large radii, and only close to the current resolution of the observations do the contributions of the black hole and the gas become roughly equal, though even together they do not account for the observed rotation curve. At large radii, the molecular gas can explain the observed rotation curve and is even slightly over--predicting the curve (although this is well within the uncertainties both of the unconstrained inclination and the assumed continuum--to--gas mass conversion).

\newpage
\subsection{Mass contributions}
\label{sec:mass_contr}

\begin{figure}
\begin{center}
\includegraphics[width=8.5cm]{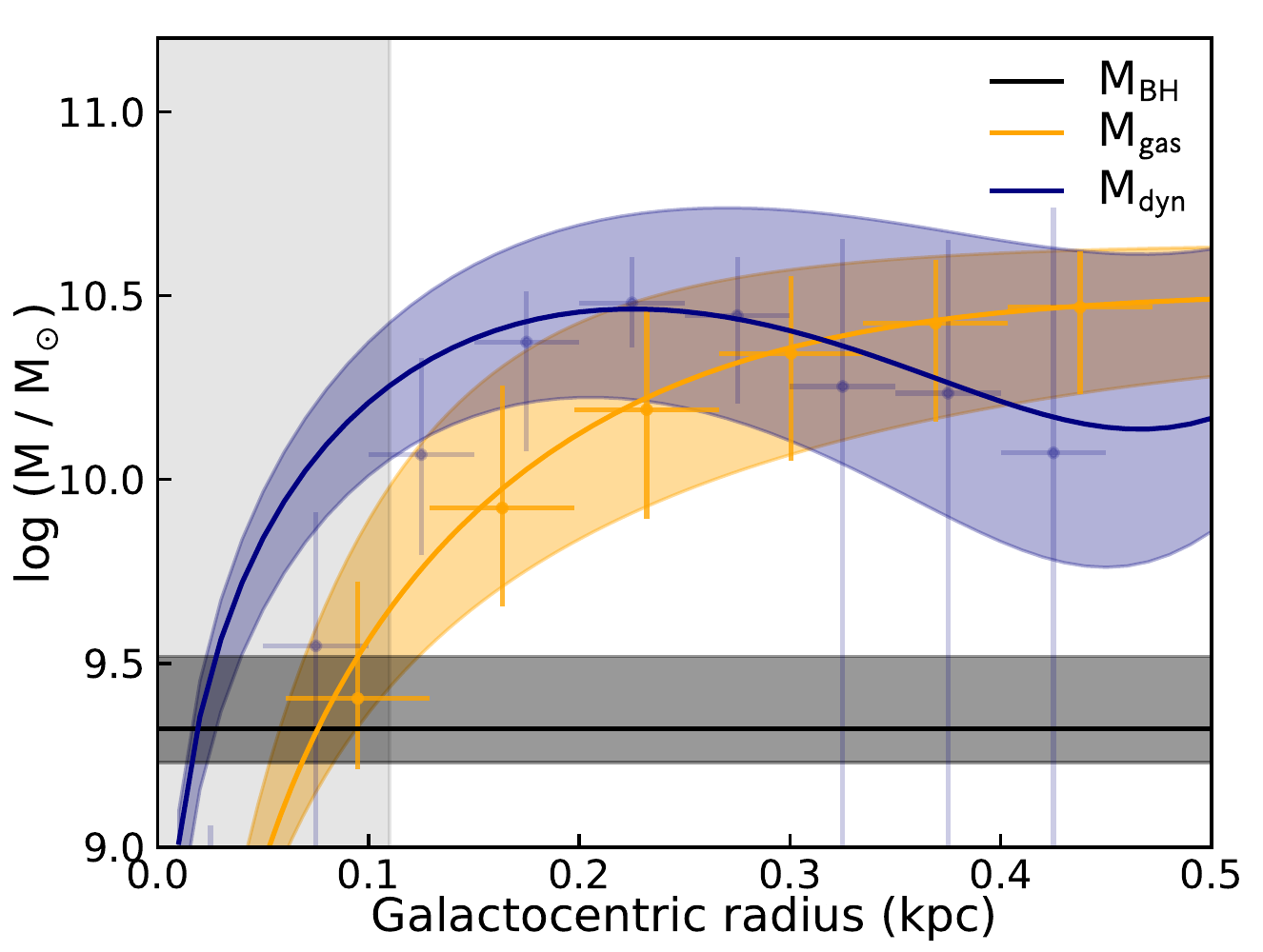}
\end{center}
\caption{Enclosed mass as a function of radius for different mass constituents. The enclosed gas mass (orange data points) is calculated by assuming the gas is traced by the dust continuum emission, where we have corrected for the effect of the beam (solid orange line). The uncertainty in this measurement (orange--shaded region) takes into account the range of allowed temperatures at each radius and uncertainties in the dust--to-gas ratio (Sec.~\ref{sec:radial}). The dynamical mass is determined from the rotation curve of the \cii\ emission line (Fig.~\ref{fig:rotcurve}) assuming the gas is gravitationally supported (violet measurements). The black hole mass measurement and uncertainties are shown in black. The resolution of the observations is marked by the gray--shaded region. At all observable radii, the mass of the gas is comparable or larger than the mass of the black hole.
\label{fig:mass}}
\end{figure}

Having obtained a rotation curve based on the \cii\ emission in Sec.~\ref{sec:decomposition}, we can convert this curve into an estimate of the enclosed dynamical mass under the assumption that the galaxy is gravitationally supported. The enclosed dynamical mass as a function of radius is shown in Fig.~\ref{fig:mass}. Here the black hole mass measurement is discussed in Sec.~\ref{sec:intro}, and the gas mass measurement is discussed in Sec.~\ref{sec:decomposition}. One interesting point is that most of the mass is very centrally located, within the inner region with a radius of 110\,pc. Beyond this radius, the dynamical mass is roughly constant, suggesting that the host galaxy of J2348--3054 is very compact.

In this mass analysis, we have so far ignored the contribution of stars. We see that at roughly the size of the beam ($\sim$100\,pc), there is a slight discrepancy between the dynamical mass estimate and the combined mass of the black hole and the molecular gas. This discrepancy could be alleviated if there was a very centrally located stellar component with a mass comparable to the gas mass ($(4\,\pm\,2)\times 10^{9}$\,M$_\odot$). The broad agreement between the dynamical mass constraints and the gas mass measurements at large radii, however, suggests that the stellar mass contributes little at larger radii. Such a compact stellar component (i.e., large bulge--to--total stellar mass ratio) is also predicted by recent simulations \citep{marshall20}, and would make it challenging to detect the stellar light from quasar host galaxies like J2348--3054 using near-infrared telescopes such as the James Webb Space Telescope. We note that the observation of a compact stellar component together with a disk of cold gas is qualitatively similar to predictions from zoom--in simulations of $z \sim 7$ quasar host galaxies \citep{lupi19}, and the increased total gas mass fraction in this galaxy is consistent with trends seen previously of increasing gas fractions with redshift \citep{carilli13,tacconi20,walter20}.

\subsection{Black hole sphere of influence}
\label{sec:BHSoI}
One particularly interesting measurement is trying to resolve the region where the black hole dominates the gravitational potential, the so--called black hole sphere of influence. Such observations with ALMA are becoming routine for nearby galaxies \citep[see][and references therein]{cohn21}. Resolving the black hole sphere of influence would allow us to directly measure the kinematic effects of the black hole, providing a dynamical constraint on the mass of the black hole. This would be of particular importance for a system at $z \sim 7$, providing a calibration--point for the relationship between UV--line widths and black hole mass for the first quasars in the Universe.

We will estimate the radius of the black hole sphere of influence as the radius where the enclosed mass in stars, gas, etc. becomes comparable to the mass of the black hole. We can see from Fig.~\ref{fig:mass} that at the current resolution of 110\,pc the gas mass alone is greater than the black hole mass. If we extrapolate the molecular gas curve to smaller radii, we find that at $75 \pm 20$\,pc the molecular gas mass becomes smaller than the black hole mass. This is the maximum radius for the black hole sphere of influence, as it ignores any stellar mass contributions, which could be important at small radii. If we instead take the radius where the black hole mass is half of the dynamical mass, we find a black hole sphere of influence radius of $35 \pm 10$\,pc. Both estimates are well below the current resolution of our observations.

\subsection{Toomre--Q parameter}
\label{sec:toomreQ}

In our resolved observations we can start looking at the stability of the gas against gravitational perturbations. For the gas in differentially rotating disk galaxies this can be represented by the so-called Toomre--$Q$ parameter, where $Q = \sqrt{2} \sigma_v v_{\rm rot} / \pi G r \Sigma_{\rm gas}$ \citep{toomre64,goldreich65}. In this equation, $\sigma_v$ is the dispersion of the gas, $v_{\rm rot}$ its rotational velocity, and $\Sigma_{\rm gas}$ the surface density at a radius $r$. Toomre--$Q$ values below one indicate regions of gas that are unstable to gravitational collapse and can therefore form stars, whereas Toomre--$Q$ values much greater than one indicate gas that is stable against gravitational collapse.

We measure the radial profile of the Toomre--$Q$ parameter in our data by computing the surface mass density of the molecular gas from the continuum observations where we correct for the effect of the beam (see Sec.~\ref{sec:decomposition}). We further can measure the beam--corrected rotational velocity profile and velocity dispersion profile directly from the kinematic modeling (see Sec.~\ref{sec:decomposition}). We find that these radially averaged Toomre--$Q$ parameters are consistent with unity for the full range of radii covered by our observations. We note that in the above calculations we only take into account the contribution of gas. The addition of stars would further lower the total Toomre--$Q$ parameter. This indicates that the gas disk is likely gravitationally unstable and can form stars, consistent with the observed high star formation rate of the galaxy.

\subsection{Scale height of the molecular gas}
\label{sec:scale_height}
The total molecular gas mass enclosed in the central 110\,pc is 4$\times$10$^{9}$ M$_\odot$ (Sec.~\ref{sec:beam}). Assuming a simple spherical geometry in the very centre, the average volume density is thus $\rho$=(3$\times$10$^{9}$ M$_\odot$)/(4/3\,$\pi$\,(110\,pc)$^3)$=3.6$\times$10$^{-20}$\,g\,cm$^{-3}$, or an H$_2$ number density of $n_{\rm H2}$=1.1$\times$10$^4$\,{\rm cm}$^{-3}$. We note that this is close to the canonical volume density of the centers of giant molecular clouds ($n_{\rm H2}\sim10^4$\,cm$^{-3}$, see e.g.{} \citealt{lada10}). We now take the central surface density of $\Sigma_{\rm H2}\sim10^{5}$\,M$_\odot$\,pc$^{-2}$ (Sec.~\ref{sec:beam}) and convert it into a molecular hydrogen column density of $N_{\rm H2}=6.3\times10^{24}$\,cm$^{-2}$. Using the volume density derived above, this leads to a total thickness of the molecular gas disk of $h\,=\,5.7\times10^{20}$\,cm or $\sim$\,190\,pc. We here assume half of this value (95\,pc) as the scale height of the gas.

We can also estimate of the height of the molecular gas by assuming hydrostatic equilibrium, which connects the scale height of the molecular gas, $h$, to the velocity dispersion of the gas, $\sigma_{\rm gas}$, and the mid--plane density of the gas $\rho$ (derived above) using the following equation \citep{vanderkruit81}:
\begin{equation}\label{eq_vanderkruit}
h=\frac{\sigma_{\rm gas}}{\sqrt{2 \pi G \rho}}.
\end{equation}
For an average velocity dispersion of 160\,km\,s$^{-1}$ (Fig.~\ref{fig:moments} and Tab.~\ref{tab:modelpar}) we derive a scale height of the molecular gas of $\sim$40\,pc. 
Considering that these are back--of--the--envelope calculations, both methods give very similar scale heights, leading to an overall thin disk with an oblateness/flattening of order $\sim$\,100\,pc/1000\,pc\,$\sim$\,0.1\footnote{We acknowledge that such a `flat' geometry would imply higher volume densities than derived above (where spherical symmetry was assumed).}.

\section{Discussion and Summary}
\label{summary}
We present ALMA $\sim$\,200\,pc imaging of the \cii{} line and the underlying dust continuum of the $z\,=\,6.9$ quasar J2348--3054, the highest--angular resolution observations yet obtained for a distant quasar host galaxy. The observations reveal very compact dust continuum and \cii{} emission, reaching extreme densities in the very central region. We derive a minimum dust temperature of 132\,K for the central resolution element, which leads to a very high TIR luminosity in that region. Converting this luminosity to a star formation rate, using standard assumptions that the dust is heated by star formation, leads to an extremely high central SFR and, correspondingly, SFR surface densities ($>$\,10$^4$\,M$_\odot$\,yr$^{-1}$\,kpc$^{-2}$, Sec.~\ref{sec:beam}). 
Such high densities could only be measured due to the very high resolution reached in the present observations. Similar high--resolution observations of a larger sample are needed to investigate if such densities are a common property of the most distant quasar host galaxy population. Such observations would also help constrain the contribution of the SMBH to the heating heating of the dust in the central $\sim$\,100\ parsecs of quasars.

The total gas mass in the central 200\,pc beam is $M_{\rm H2}$=$(4\pm2)\times 10^{9}$\,M$_\odot$, or about a factor of two higher than that of the central supermassive black hole. Therefore, the gas kinematics in the center are not dominated by the influence of the supermassive black hole. Converting the above gas mass to an H$_2$ column density yields densities well within the Compton--thick regime ($\rm{N}_{\rm{H}} > 10^{24}\ \rm{cm}^{-2}$). Such high column densities should imply that the quasar is heavily obscured, consistent with the recent non-detection with \textit{Chandra} \citep{wang21b}. However, rest--frame UV spectra \citep{venemans13, wang21b} do not present particular reddening / extinction (a situation often found in type~1 quasars). This suggests that most of the quasar emission can escape either through lower density pockets of gas, which is a likely possibility since we are observing this galaxy nearly face--on, or the quasar is slightly offset from the central Compton--thick gas. Unfortunately, the astrometric uncertainties for the GAIA--corrected optical position of the quasar \citep{venemans20} are too large to explore the latter possibility.

We find that the interstellar medium, as traced by dust and \cii{} is smooth out to radii of $\sim$\,500\,pc, and its kinematics are consistent with a simple `flat' rotation curve. Despite the regular velocity field, the gas has a significant velocity dispersion, with an average value of $\sim$\,160\,km\,s$^{-1}$. Assuming hydrostatic equilibrium, this leads to a `puffed--up' disk with an oblateness of $\sim\,0.1$, still consistent with a rather thin disk. It should be noted, however, that other kinematic models (such as solid body rotation or a Keplerian decline) cannot be ruled out with the available data (Sec.~\ref{sec:kinematics}, Appendix~\ref{sec:model_pars}). No evidence for outflows is found in the data.

Evidence is building for a dramatic change in the host galaxies of supermassive black holes with redshift. At low redshift, it is well known that extreme supermassive black holes ($\rm M_{BH} \ge 10^{10}$), are always hosted by large elliptical galaxies \citep{kormendy13}. At high redshift, there is growing evidence for disk host galaxies, even for the most extreme black holes \citep[e.g.,][]{neeleman21}. J2348--3054 represents the most distant such example to date, as well as the clearest example based on resolved galaxy dynamics. While the change from disk to elliptical host galaxies with increasing cosmic time is consistent with the general scenario of mergers of disk galaxies leading to elliptical galaxies, the details of the demographics, and the implications for the evolution of SMBHs and their host galaxies, remains to be determined. 

A second result from our dynamical study is the conclusion that the dynamical mass of J2348--3054 enclosed within the largest observed radius can be almost completely explained by the gas mass within the uncertainties, i.e., there is no need for stars (or dark matter). While the limits are not highly constraining, the data imply a system in which the gas mass fraction $\rm M_{gas}$/M$_{\rm stars}\,>\,$1. This result is consistent with observations of main sequence star forming galaxies, which show a change in the cool gas to stellar mass fraction from $\le 0.1$ in the nearby Universe, to $\sim 1$ at $z\,>\,2$ \citep{tacconi20, walter20, aravena20}.

A major goal still remains for J2348--3054: a direct measurement of the black hole mass, using gas dynamics. However, the very compact gas and dust distribution (and hence compact mass distribution), makes such an observation very challenging in the case of J2348--3054. The kinematic signature of a $\sim\,2\,\times$10$^{9}$\,M$_\odot$ black hole will be evident at a radius where this mass dominates, i.e. where the mass contribution of the other baryons in the host galaxy will be of order $\sim$10$^{9}$\,M$_\odot$. Our estimates based on the current observations suggest that this happens at a resolution of about 50\,pc (Sec.~\ref{sec:BHSoI}). Such a resolution can just be reached with ALMA in the most extended configuration (leading to 18 mas at the observed \cii{} frequency of $\simeq$240 GHz). This is a beam area that would be 4 times smaller than the current resolution. If the \cii\ flux was distributed uniformly, this would imply a four times smaller flux per beam, which would be hard to detect with integration times $<$\,100\,h. If however the emission were to peak further towards the center, the flux in the central beam(s) may be sufficiently high to measure the black hole sphere of influence.

\appendix

\begin{figure*}
\centering
\includegraphics[width=0.55\textwidth]{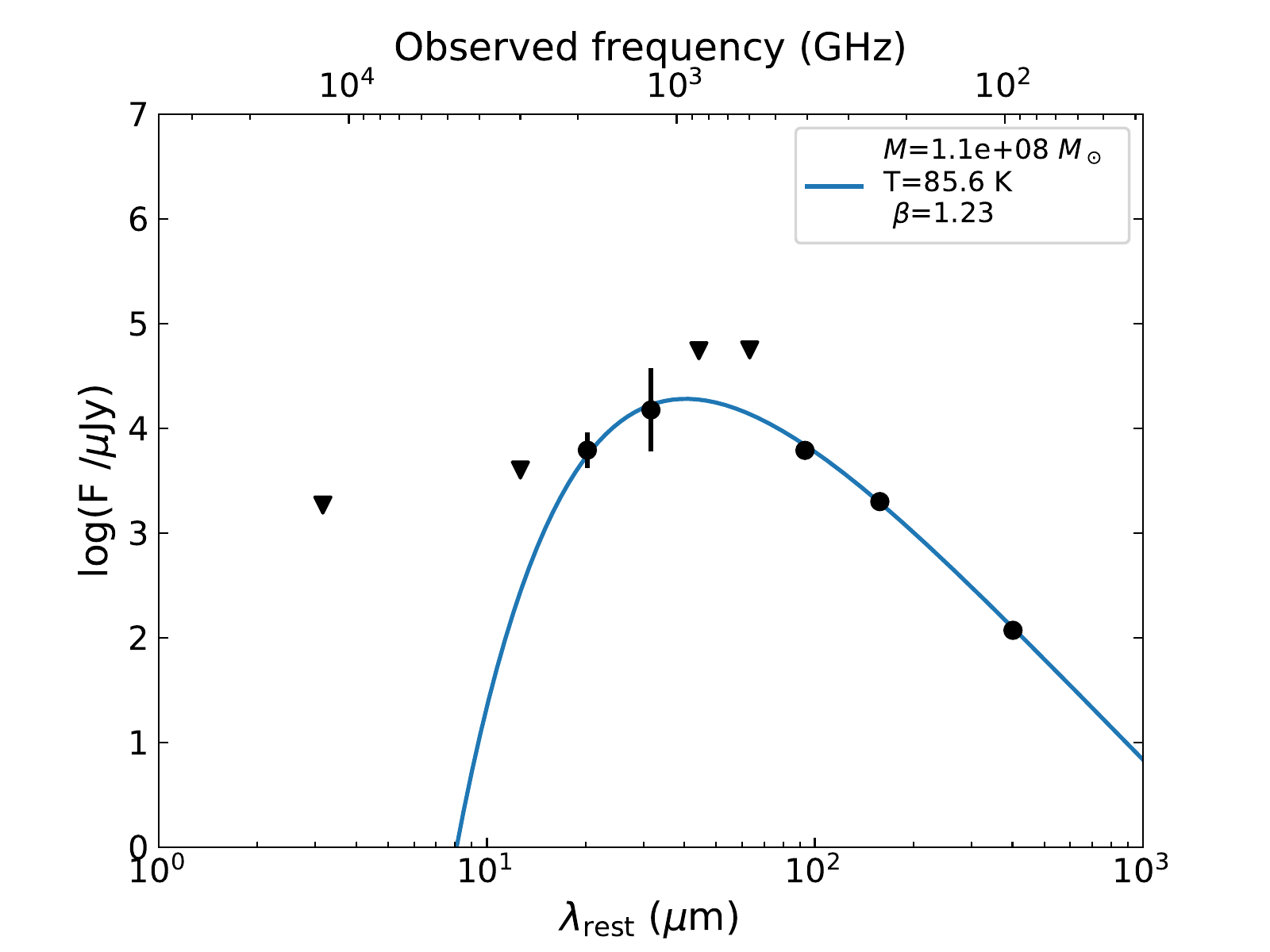}
\includegraphics[width=0.44\textwidth]{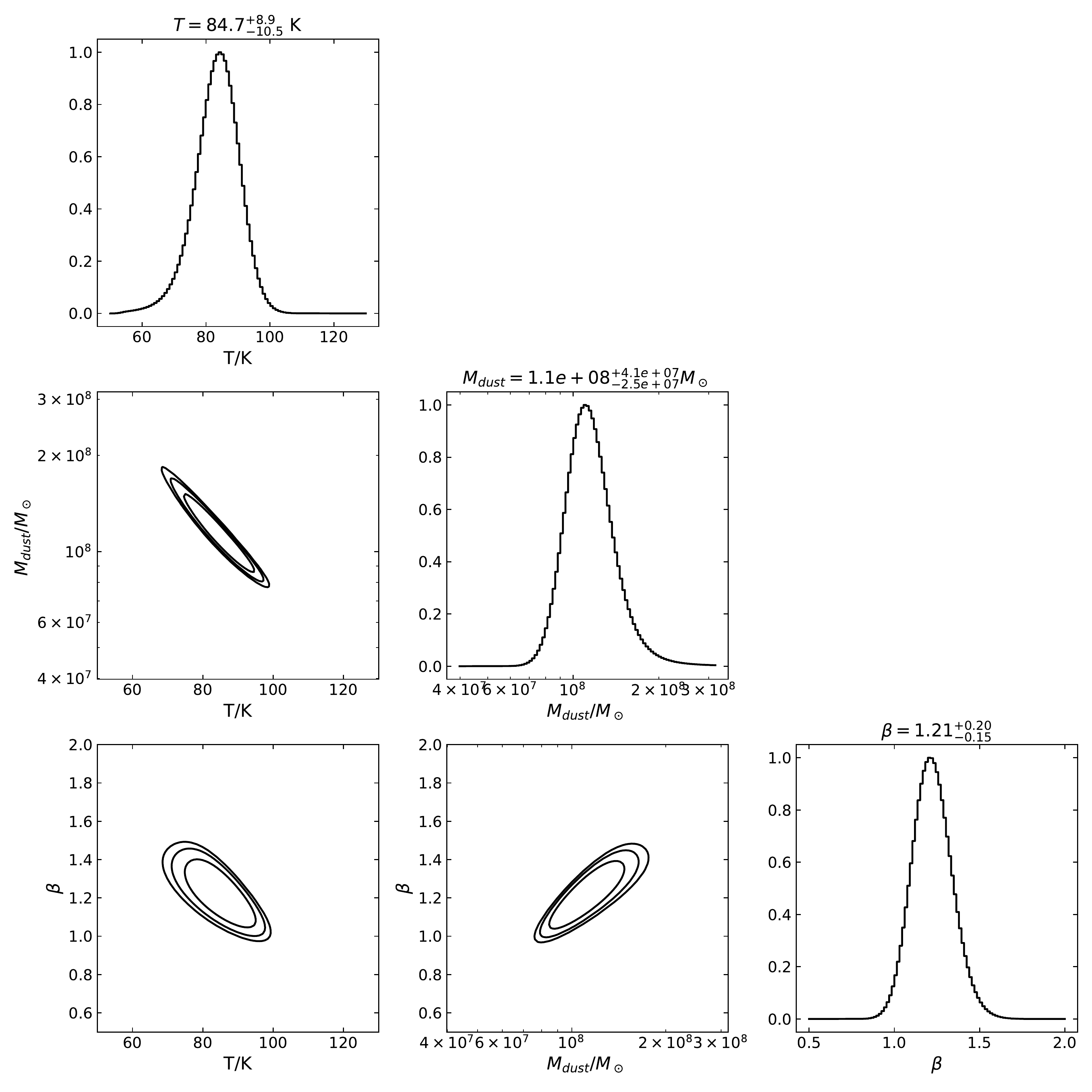}
\caption{ {\em Left:} Dust spectral energy distribution of J2348–-3054 based on the data presented in Tab.~\ref{tab:SED}. The best--fit dust SED for the long wavelengths is shown as a blue line and the best--fit parameters are given in the inset. {\rm Right:} Posterior distributions of the dust SED parameters with $1,2$ and $3\sigma$ contours.
\label{fig:SED}}
\end{figure*}

\section{Channel maps}
\label{sec:channels}

\begin{figure*}[b]
\centering
\includegraphics[width=0.95\textwidth]{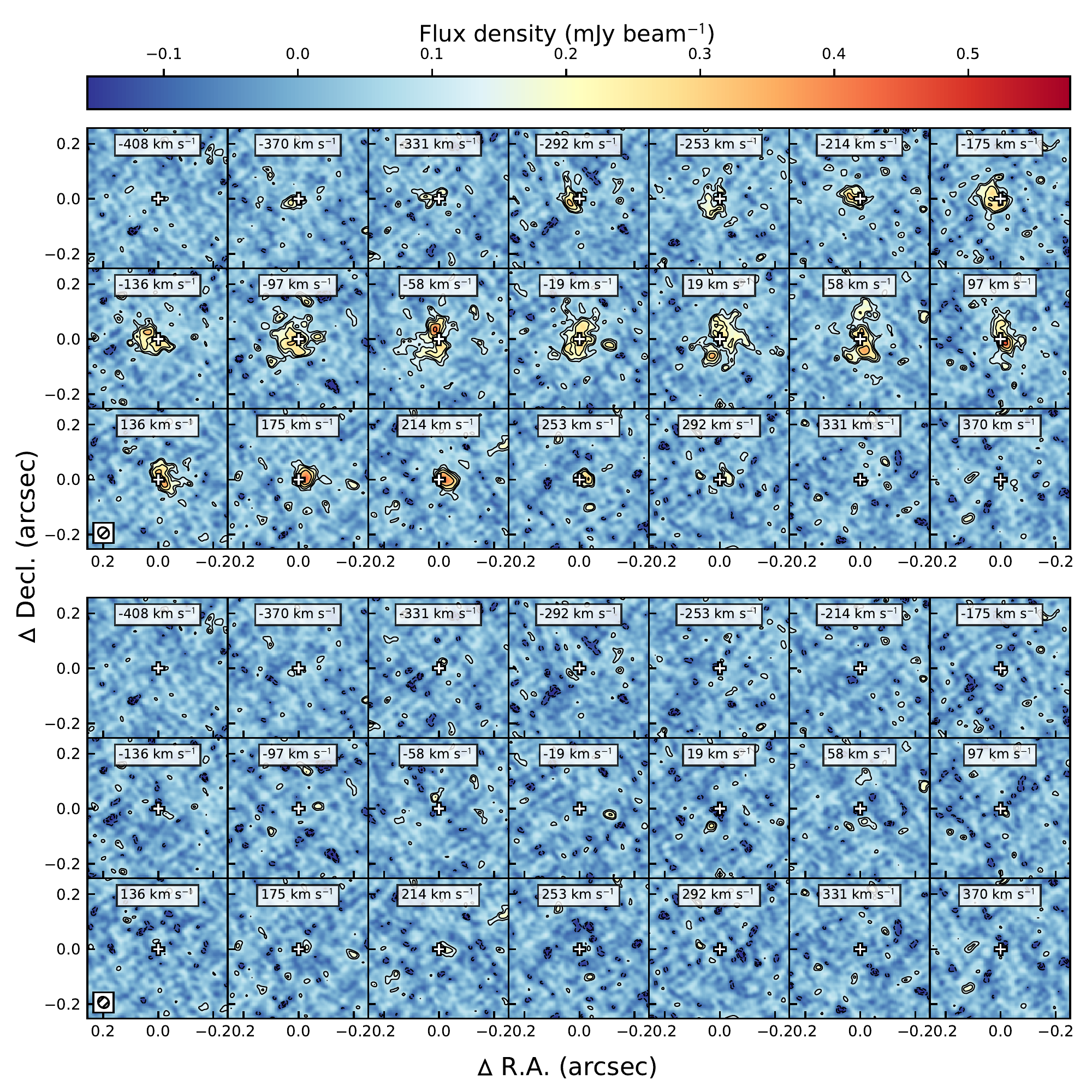}
\caption{
\emph{Top panels:} \cii\ channel maps of J2348–-3054 with a channel width of 31.2\,MHz (38\,km\,s$^{-1}$). Contours start at $\pm\,2\,\sigma$ and increase in powers of $\sqrt{2}$, where $\sigma = 53~\mu$Jy~beam$^{-1}$. Positive flux is shown as full contours, negative emission is shown as dashed contours. The plus sign marks the position of the dynamical center of the \cii\ emission as determined from the kinematic modeling. Velocities are relative to $z\,=\,6.9018$. The beam is shown as an inset in the bottom left panel. \emph{Bottom panels:} \cii\ channel maps of J2348--3054 after subtraction of the infinitely thin disk model with constant rotational velocity. Little substructure can be seen in the individual channels indicating that the model provides an accurate fit of the data although this is a common feature among all of the models that were tested.
\label{fig:channels}}
\end{figure*}

Fig.~\ref{fig:channels} shows the \cii{} channel maps of  J2348--3054 (after continuum subtraction) at a resolution of 0.035$''$ ($\sim$200\,pc). As the velocity increases, the  \cii{} emission is shifting from west to east, indicative of rotation. In the bottom panels of Fig.~\ref{fig:channels} we show the residual \cii{}  channel  maps  after  subtraction  of  the  infinitely  thin  disk  model  with  constant rotational  velocity (Sec.~\ref{sec:modeling}, first column in Tab.~\ref{tab:modelpar}). Little substructure  is present in the individual residual channels, indicating that the model provides an acceptable fit of the data. We note that all other models described in Sec. \ref{sec:model_pars} provide a similar fit to the data. This indicates that with the current sensitivity we cannot use the dynamics of the \cii\ emission line to distinguish between different model rotation curves.

\section{Dust Spectral Energy Distribution}
\label{app:band_8}
We have obtained band~8 observations of J2348--3054 using the ALMA Compact Array (ACA) to secure a measurement of the dust continuum at 406.865 GHz. Observations were carried out on 2021, August 10 and 13 as part of the program 2019.2.00053.S. These observations resulted in a beam size of $4.6''\times 2.7''$. A 2D Gaussian fit of the source shows that the quasar host is unresolved, and has a continuum flux density of $6.17 \pm 0.63$ mJy (S/N\,=\,9.8). 

We have compiled all relevant dust SED data for J2348--3054 from the literature (and data archives) and have show the results in Tab.~\ref{tab:SED}. Fluxes were extracted from the {\em Herschel} SPIRE and PACS cleaned maps using Source Extractor in fixed apertures of 10'' in radius. Forced photometry in equal apertures was conducted in surrounding blank regions to estimate the uncertainties. Significant detections are seen in the SPIRE $250\,\mu$m and PACS $160\,\mu$m channels. We use a $3\sigma$ non-detection flux limit for the other channels, except in PACS $3000$\,GHz ($100\,\mu$m). A marginal $\sim 3\sigma$ detection is obtained in this  channel, but the PSF indicates this is likely a noise fluctuation; we use the $5\sigma$ limiting flux in this channel.

We fit the photometry with a dust SED of the form $F_\nu \propto M_{\rm{dust}} (\nu/\nu_0)^\beta [B(\nu,T) - B_{\rm{CMB}}(\nu)]$ where $B(\nu,T)$ corresponds to black-body radiation and $B_{\rm{CMB}}$ is emission from the CMB, subtracted to obtain the brightness contrast. The fit likelihood is calculated on a three-dimensional grid, with no priors for the three parameters $(T, M_{\rm{dust}}, \beta)$. We show the resulting best--fit dust SED and the accompanying corner plots in Fig.~\ref{fig:SED}. The best solution has $\chi^2 = 2.35$ for $2$ effective degrees of freedom, indicating an excellent fit. We obtain marginalized values and $68\%$ credible intervals of $T=84.7_{-10.5}^{+8.9}$ K, 
$M_{\rm{dust}} = 11.0_{-2.5}^{+4.1} \times 10^7 M_\odot$, 
and $\beta=1.21_{-0.35}^{+0.20}$. The corresponding total infrared luminosity is $L_{\rm TIR} = 3.2 \times 10^{13}$ L$_\odot$.

Note that we use the optically--thin approximation in this overall SED fit because the dependence of the area of the emitting region, $A$, on wavelength, is unknown (see Section \ref{sec:tau}). Indeed, as shown above, J2348--3054 shows strong evidence for a radially-dependent dust temperature profile (Sec.~\ref{sec:radial}). Warmer dust located nearer the center of the galaxy is expected to dominate the emission at higher frequencies, leading to a smaller $A$ and less optical depth. Such a frequency dependence is non--trivial, and beyond the scope of what can be constrained with using $5$ photometric measurements. Regardless of these considerations, the {\textit{Herschel}} photometry rules out the presence of hot gas ($T\gtrsim100$\,K) in significant quantities beyond the central region, since the peak frequency of the SED is well-constrained.

\begin{table}
   \centering
   \begin{tabular}{lcccl}
   Instrument & Freq. (GHz) & S$_\nu$ (mJy) & rms. & Source \\
   \hline
   \hline
ALMA & $94.5$ & $0.118$ & $0.013$ & \citet{venemans17} \\
ALMA & $240.575$ & $2.00$ & $0.07$ & this work \\
ACA & $406.88$ & $6.17$ & $0.63$ &  this work  \\
{\em Herschel} SPIRE & $600.0$ & $<56$ & & archival, P.I.~McMahon \\
& $856.55$ & $<55$ & &  archival, P.I.~McMahon\\
& $1200$ & $15$ & $6.0$ &  archival, P.I.~McMahon \\
{\em Herschel} PACS & $1873$ & $6.2$ & $2.0$ &  archival, P.I.~McMahon\\
&  $3000$ & $<4.0 \ (3.3)$ & $(0.8)$ &  archival, P.I.~McMahon\\
{\em WISE} W4 & $12000$ & $<1.85$ & & \citet{Wright10,Cutri14} \\
   \hline
   \hline
   \end{tabular}
   \caption{Column 1: instrument. Column 2: observed frequency (approximate central frequencies are given form Herschel and WISE bands). Column 3: flux denisty at the given frequency. Limits are at the $3\sigma$ level, except PACS $3000$ GHz which is quoted at $5\sigma$, given the source confusion in the field. Colum 4: noise at the given frequency. Column 5: Reference, for the Herschel observations, the PID is `{OT2\_rmcmahon\_1}'.}
   \label{tab:SED}
\end{table}

\section{Additional Qubefit Models}
\label{sec:model_pars}

\begin{figure*}
\begin{center}
\includegraphics[width=0.48\textwidth]{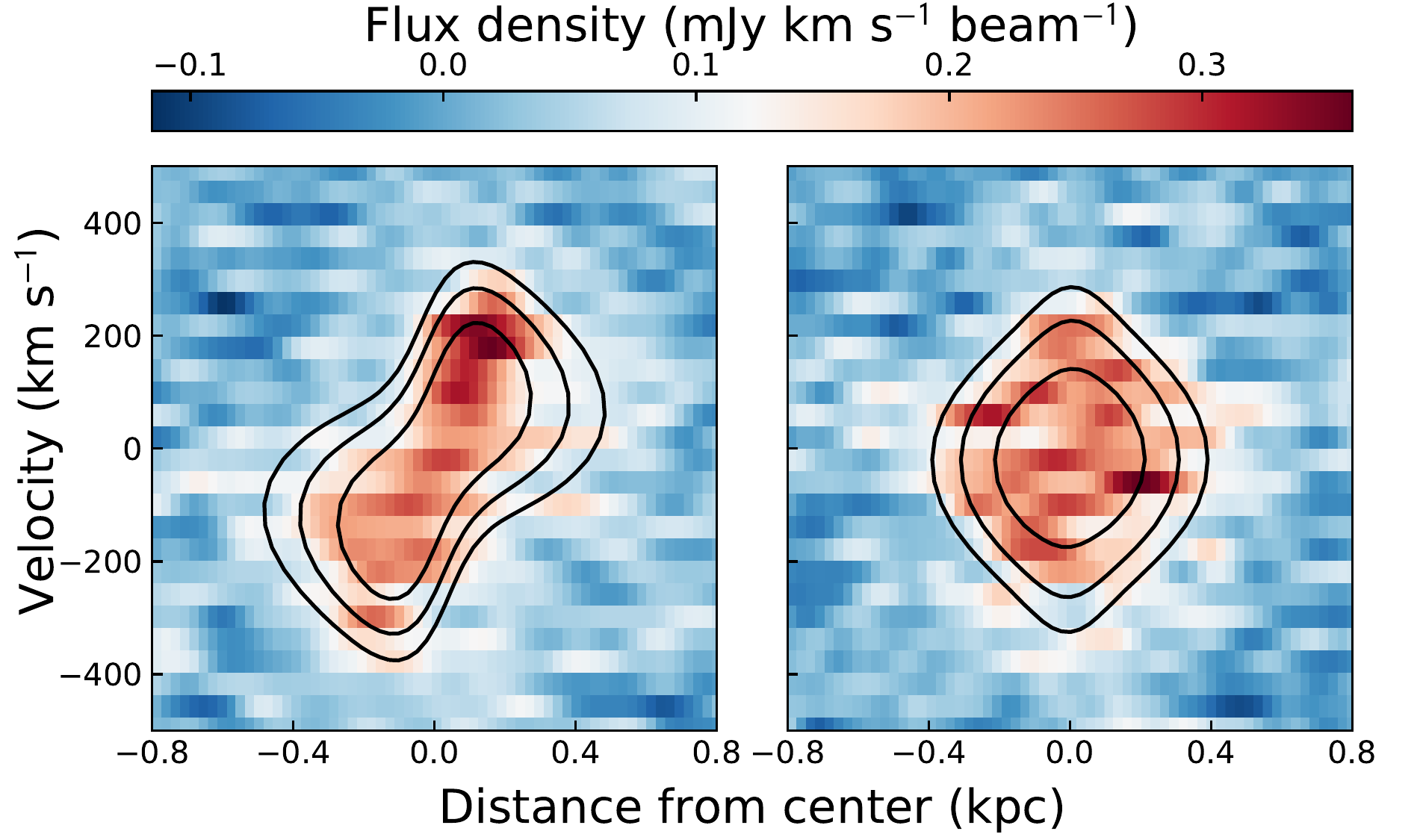}
\includegraphics[width=0.48\textwidth]{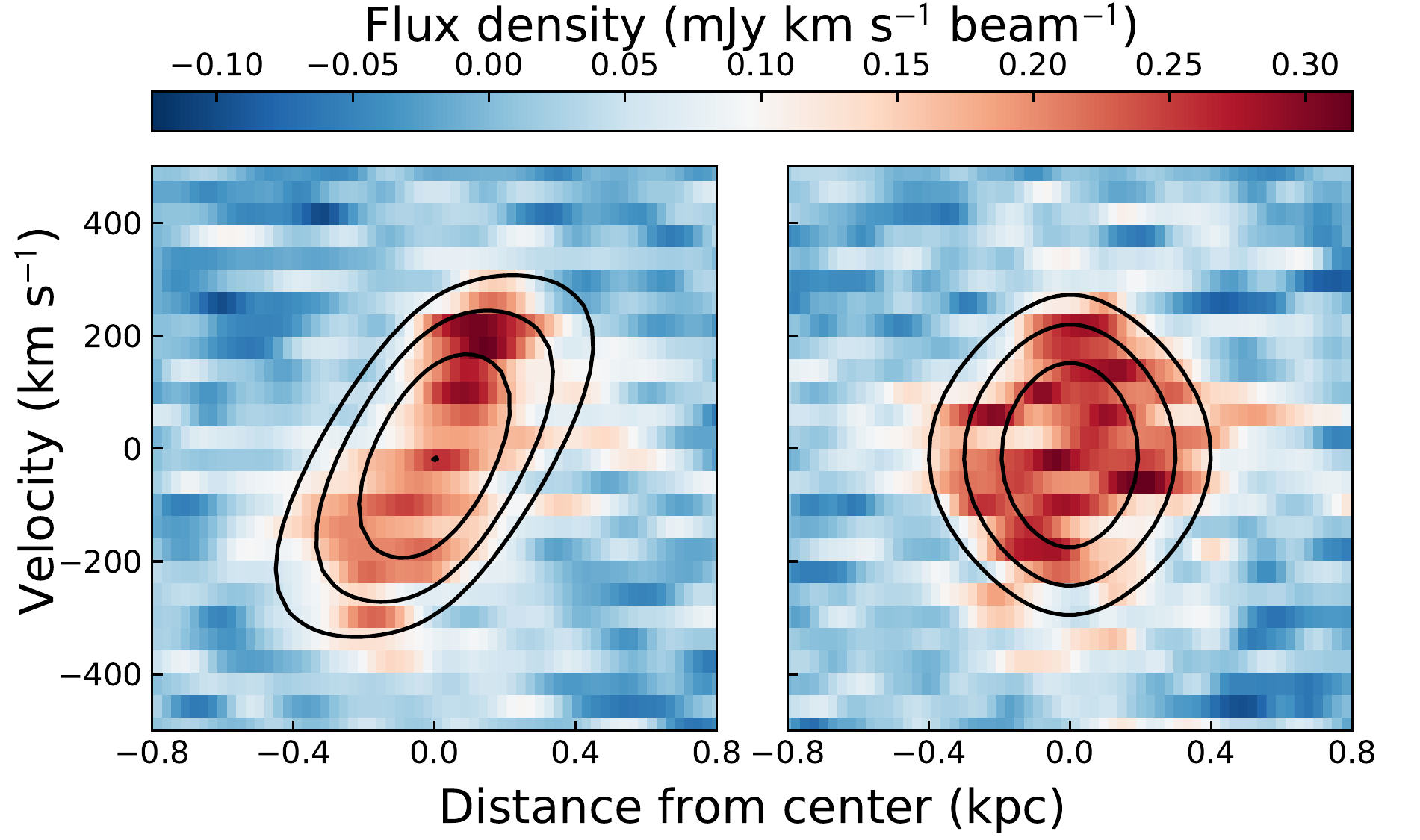}
\end{center}
\caption{Position--velocity diagrams for two additional models ({\em left:} Keplarian, {\em right:} solid body rotation). Fitting parameters are given in Tab.~\ref{tab:modelpar}.
\label{fig:pv_models}}
\end{figure*}

In Tab.~\ref{tab:modelpar} we give the model parameters of the kinematical modeling using \emph{Qubefit} \citep{neeleman21}, as presented in Sec.~\ref{sec:modeling}. Fig.~\ref{fig:pv_models} shows position velocity diagrams for two additional models (solid body and Keplarian rotation curves), analogous to Fig.~\ref{fig:pV}. These two idealized models were chosen because they bracket the possible shapes of the rotation curve, i.e., the solid body rotation curve is linearly increasing with radius whereas the Keplerian rotation curve decreases exponentially with radius. For the Keplerian curve we fix the curve to a velocity of 101.6\,km\,s$^{-1}$ at 1\,kpc, which corresponds to the rotational velocity of a point source with a mass of $2.1 \times 10^9\,$M$_\odot$. The good agreement we see between these idealized models and the data shows that despite the high resolution observations, the data cannot distinguish between different types of rotation curves. We therefore adapt the flat rotation curve as our fiducial model.

\begin{table*}
\centering
\caption{Model parameters
\label{tab:modelpar}}
\begin{tabular}{lllll}
\hline
& & Flat & Solid body & Keplerian\\
\hline
R. A. & (J2000) & 23:48:33.34541(9) & 23:48:33.34530(8) & 23:48:33.34545(5)\\
Decl. & (J2000) & -30:54:10.2963(8) & -30:54:10.2964(11) & -30:54:10.2969(12)\\
$z$ & & 6.90131(12) & 6.90144(13) & 6.90123(9)\\
$\alpha$ & ($^\circ$) & $275.3 \pm 1.9$ & $269.4 \pm 2.6$ & $279.8 \pm 2.1$\\
$i$ & ($^\circ$) & $<25.7^a$ & $<24.1^a$ & $38.8 \pm 1.3$\\
$I_0$ & (mJy kpc$^{-2}$) & $11.6 \pm 0.4$ & $9.1 \pm 0.3$ & $12.3 \pm 0.4$\\
$R_d$ & (kpc) & $0.254 \pm 0.007$ & $0.294 \pm 0.010$ & $0.275 \pm 0.008$\\
$v_{\rm rot}$ & (km s$^{-1}$) & $>375^a$ & $>368^{a,b}$ & $101.6^c$\\
$\sigma_v$ & (km s$^{-1}$) & $161 \pm 4$ & $190 \pm 5$ & $157 \pm 4$\\
\hline
\multicolumn{5}{l}{$^a$ 3$\sigma$ limits}\\
\multicolumn{5}{l}{$^b$ rotational velocity at $R_d = 0.3$~kpc}\\
\multicolumn{5}{l}{$^c$ velocity at 1\,kpc for a black hole with a mass of $2.1 \times 10^9$\,M$_\odot$}\\
\end{tabular}
\end{table*}

\acknowledgments

We thank the referee for a constructive report that helped to improve the paper. FW, MN, BV, RAM and SB acknowledge support from the ERC Advanced Grant 740246 (Cosmic\_Gas). We thank Mladen Novak for helping with the flux determinations presented in this paper. This paper makes use of the following ALMA data: ADS/JAO.ALMA\#2018.1.00012.S and ADS/JAO.ALMA\#2019.2.00053.S. ALMA is a partnership of ESO (representing its member states), NSF (USA) and NINS (Japan), together with NRC (Canada), NSC and ASIAA (Taiwan), and KASI (Republic of Korea), in cooperation with the Republic of Chile. The Joint ALMA Observatory is operated by ESO, AUI/NRAO and NAOJ. The National Radio Astronomy Observatory is a facility of the National Science Foundation operated under cooperative agreement by Associated Universities, Inc.

\vspace{5mm}
\facilities{ALMA, Herschel, WISE}

\software{CASA \citep{mcmullin07}, Qubefit \citep{qubefit21}, RADEX \citep{vandertak07}, Interferopy \citep{interferopy}}

\facility{ALMA}

\end{document}